# Deformation and Stress Evolution during Laser Powder Bed Fusion of Semi-Crystalline Polyamide-12

Zhongfeng Xu[1], Wei Zhu[2]*, Lionel Freire[1], Noëlle Billon[1], Jean-Luc Bouvard[1]*, Yancheng Zhang[1]*

*1 MINES Paris, PSL University, Centre for material forming (CEMEF), UMR CNRS, 06904 Sophia Antipolis, France*

*2 College of Mechanical and Vehicle Engineering, Hunan University, 410082, Changsha, China*

Corresponding author: Wei Zhu (zhuwei@hnu.edu.cn); Jean-Luc Bouvard (jean-luc.bouvard@minesparis.psl.eu); Yancheng Zhang (yancheng.zhang@minesparis.psl.eu);

## Abstract

Laser powder bed fusion (L-PBF) of semi-crystalline polymers such as polyamide-12 (PA12) has found increasing use in various industrial applications. However, achieving high dimensional accuracy remains a significant challenge. Despite the seemingly straightforward layer-by-layer manufacturing concept, the L-PBF process involves complex thermal histories and strongly coupled multiphysics, making the evolution of stress and deformation mechanisms still not fully understood. To address this, a comprehensive three-dimensional thermo-mechanical modeling framework is developed to simulate the L-PBF process of PA12. The model for the first time incorporates transient heat transfer, phase transformation induced volumetric shrinkage, thermo-viscoelasticity, and a modified non-isothermal crystallization kinetics. To alleviate the computational burden of part-scale simulations, a dual-mesh strategy is employed to efficiently couple thermal and mechanical fields without compromising numerical accuracy, which also enables the framework to handle L-PBF simulations of arbitrarily complex three-dimensional geometries.

Particular attention is paid to the role of mechanical and thermal boundary conditions. Specifically, the underlying powder bed is modeled as a fictitious viscous medium, providing support while permitting upward displacement. Additionally, a radiative heat loss boundary condition, which more closely approximates the actual physical process, is applied to the top powder surface. The incorporation of this radiation effect significantly enhances the crystallization rate and improves the agreement with experimentally measured warpage. The model is validated against experimental warpage data under various preheating temperatures. Furthermore, strain decoupling analysis for the first time reveals that displacement induced by phase transformation is approximately 10 times greater than that caused by thermal expansion, highlighting the dominant role of crystallization-induced shrinkage in warpage formation. Numerical tests also indicate that warpage is highly sensitive to the preheating target temperature of the PA12 powder bed, while the temperature of the newly recoated powder within the tested range has a limited effect. This work provides a predictive modeling foundation for future optimization of polymer L-PBF processes at part-scale, particularly in controlling deformation and improving dimensional accuracy.

## Introduction

Laser powder bed fusion (L-PBF) has emerged as one of the most promising additive manufacturing (AM) technologies and has been successfully applied to a broad range of material systems, including metals, ceramics, and polymers [1–3]. Currently, polymer-based L-PBF is primarily employed in prototype development and non-load-bearing applications, such as brackets, medical implants, and sports equipment [4,5]. Recent advancements in the L-PBF processing of high-temperature polymers have further expanded its potential for applications in the aerospace sector [6–8].

Nevertheless, ensuring consistency and repeatability in the dimensional accuracy and mechanical performance of L-PBF manufactured parts remains a significant challenge. These limitations are primarily attributed to several intrinsic factors, including incomplete fusion, residual porosity, and heterogeneous deformation occurring during and after fabrication [5,9,10].

In recent decades, substantial progress has been made in the L-PBF of semi-crystalline polymers. Both experimental and numerical studies have advanced the understanding of the underlying multiphysics phenomena, including complex thermal histories, particle consolidation dynamics, and crystallization kinetics, all of which are critical to achieving high-quality parts. However, warpage remains one of the most persistent issues, typically occurring during or after the fabrication process. In contrast to metallic L-PBF, polymer-based processes typically do not employ a rigid build substrate, resulting in insufficient mechanical constraint. This allows for uneven accumulative shrinkage throughout the layer-wise build, leading to significant deformation and a loss of dimensional accuracy. In extreme cases, warpage may cause part detachment or collisions during powder recoating, directly resulting in print failure [11]. Therefore, mitigating warpage is essential for improving the geometric fidelity and reliability of polymer L-PBF.

Although prior studies have linked warpage primarily to non-uniform temperature distributions and complex thermal histories [12], the detailed role of transient thermal profiles in driving phase transformations, stress accumulation, and localized deformation remains insufficiently understood. Addressing this gap is vital for advancing predictive capabilities and process optimization.

A well-recognized mechanism of warpage involves the uneven and sequential shrinkage that occurs during the cooling stage. Yan et al. [13] compared the shrinkage behavior of amorphous (PS) and semi-crystalline (PA12) polymers. Their study showed that PA12-printed parts exhibited worse dimensional precision, primarily due to the pronounced volumetric contraction during crystallization. Expanding on this, Teo et al. [14] investigated the influence of cooling rate on PA12 shrinkage using thermal mechanical analysis (TMA). Their results indicated that faster cooling rates exacerbated shrinkage during crystallization, emphasizing the impact of thermal history on macroscopic deformation in semi-crystalline polymer parts.

To further explore the intrinsic warpage potential of semi-crystalline polymers, Verbelen et al. [15] studied the dilatometric behavior of polyamides as a function of temperature. As shown in Fig.1(a), the specific volume demonstrates a distinct step-change during both melting (solid line) and crystallization (dashed line), hallmarks of first-order phase transitions. These findings underscore the need to account for phase-induced volume changes in predictive modeling. Building on these, Wang et al. [16] proposed a schematic representation of the warpage mechanism in multi-layer L-PBF processing of PEEK, a high-performance semi-crystalline polymer, as shown in Fig.1(b). Their work highlighted that warpage is predominantly triggered by crystallization-induced shrinkage in the lower layers of the build, where thermal compensation from the surrounding environment (e.g., heating lamps) is insufficient. This explanation emphasizes the layer-dependent thermal and mechanical instability inherent in the L-PBF process for semi-crystalline polymers.

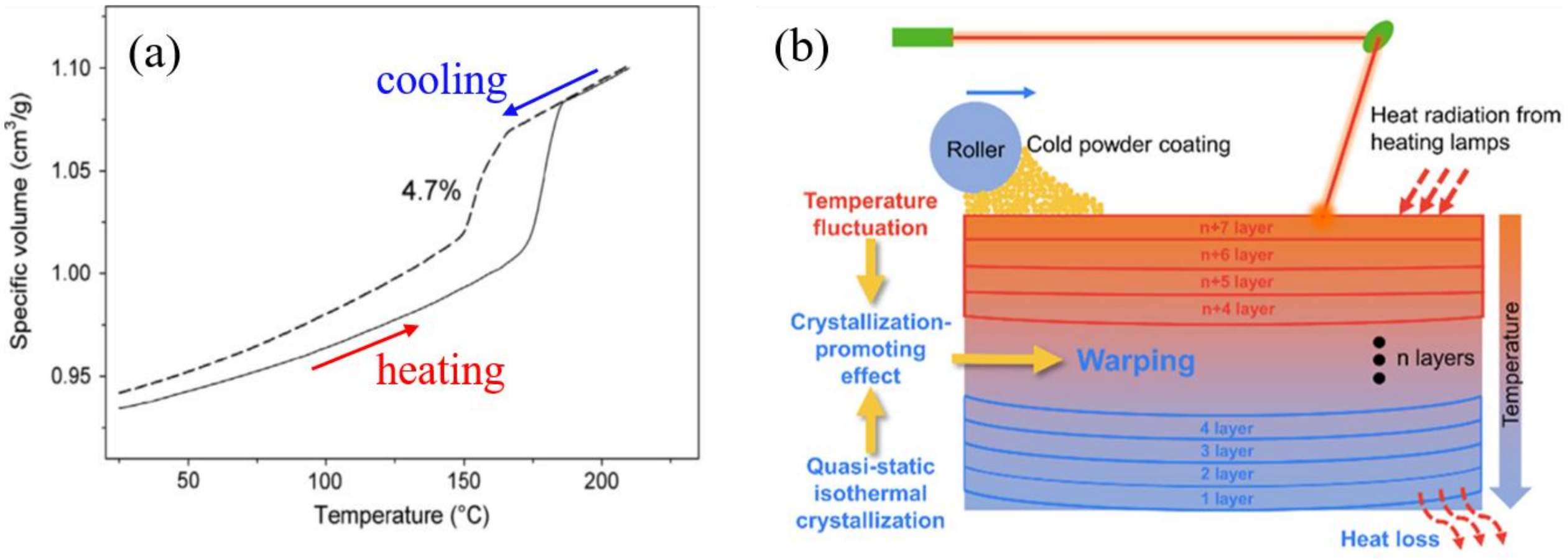

*Fig.1 (a) specific volume change of PA12 during heating and cooling process [15]; (b) the crystallization schematic in PBF LB/P process [16].*

With crystallization now widely acknowledged as a dominant source of warpage in semi-crystalline polymers, the kinetics of crystallization in materials like PA12, PP, and PEEK have been extensively studied [6,17–19]. Nonetheless, the spatiotemporal characteristics of phase transformation during L-PBF remain unclear due to a lack of in-situ diagnostic tools capable of capturing these transitions under realistic process conditions. In an effort to overcome this gap, numerous thermal numerical models have been developed to simulate temperature evolution and crystallization behavior during polymer L-PBF [20–22]. These models lay the groundwork for more advanced thermo-mechanical frameworks aimed at predicting stress development and part distortion.

For instance, Manshoori Yeganeh et al. [23] introduced a finite element (FE) model to predict distortion under varying scanning strategies, though their approach neglected the influence of surrounding powder and phase change. Shen et al. [24] improved upon this by integrating non-isothermal crystallization kinetics and relating crystallization-induced strain to volume shrinkage. However, their mesoscale model was limited in scope, as it did not capture the cumulative warping effect across multiple layers or the evolving thermal interactions between consecutive layers due to powder deposition. Amado et al. [25,26] proposed a state-of-the-art modeling approach by incorporating a thermo-viscoelastic constitutive law for polymers. They also accounted for the abrupt change in specific volume associated with crystallization. However, their framework remained two-dimensional and relied on oversimplified assumptions for the thermal history inherent to layer-by-layer fabrication.

To date, there remains a lack of comprehensive numerical frameworks for predicting displacement and stress development in polymer AM at part-scale. In parallel, another critical modeling challenge lies in the appropriate selection of boundary conditions for polymer L-PBF. In the practical application of L-PBF for polymers, the laser processed part is supported by the underlying powder bed, which limits its downward displacement while allowing a degree of upward and lateral deformation. However, many existing numerical studies either omit explicit discussion of mechanical boundary conditions or adopt oversimplified constraints such as fully fixed boundary [24,27]. Such assumptions are inadequate for accurately capturing the true deformation behavior and thus compromise the reliability of displacement and stress predictions. In addition, during the building process, heat transfer between the top surface of the part and the surrounding chamber air is often neglected in the literature [28], despite its potential impact on the multiphysics evolution and overall simulation fidelity.

This study develops a versatile part-scale thermo-mechanical finite element model for the L-PBF process of semi-crystalline polymers. The proposed framework integrates layer-by-layer powder recoating and laser scanning, crystallization-induced volumetric shrinkage, and temperature-dependent viscoelastic constitutive behavior. Leveraging this model, we first investigate the role of mechanical boundary conditions in realistically representing the support powder, aiming to achieve deformation patterns consistent with experimental observations. Next, the dominant contribution of phase transformation—particularly crystallization induced volume change—to part displacement and stress evolution is quantitatively analyzed. Furthermore, the impact of key thermal parameters, such as preheating temperature and powder surface radiation, is assessed through coupled simulations and experimental validation. Notably, our findings highlight that incorporating radiative heat transfer at the powder surface is essential to accurately predict warpage in polymer L-PBF, offering new insights for improving simulation fidelity and dimensional accuracy control.

## 2. Methodology

In this work, a comprehensive numerical framework was established based on the level-set method [29]. By employing signed distance functions associated with distinct physical fields, the model is capable of accurately delineating multiple computational domains, including air/material zones, loose powder regions, and the consolidated part. The implementation details and discussion of the macroscopic thermal modeling of the L-PBF process were presented in our previous study [30].

To enhance computational efficiency at the part scale, a layer-wise energy deposition strategy was adopted based on the energy-equivalence principle [31]. The thermal and mechanical analyses were sequentially decoupled: the transient temperature field obtained from the thermal solver was used as the driving input for the mechanical simulation, without any feedback of energy dissipation, deformation, or displacement to the heat transfer calculation. The effect of the surrounding powder was neglected, whereas the underlying powder bed was modeled as a highly viscous fluid to account for its mechanical support during processing simulation.

The finite element-based thermo-mechanical model in this work was implemented using our in-house software, CIMLIB [32]. The improved non-linear thermal solver with the sub-incremental stabilization [33] and the thermo-viscoelastic constitutive law were developed in C++ to ensure robustness and flexibility.

In this section, the numerical approach that describes the relationship between the thermal and mechanical models is first introduced in Subsection 2.1. The governing equations for the thermal and mechanical analyses are then presented in Subsections 2.2 and 2.3, respectively. Next, the couplings between latent heat effects and the phase-transformation-induced specific volume change with the modified crystallization kinetics are described. Finally, the material properties, including the thermo-viscoelastic constitutive law and basic thermal parameters, are outlined.

## 2.1 Bi-mesh method

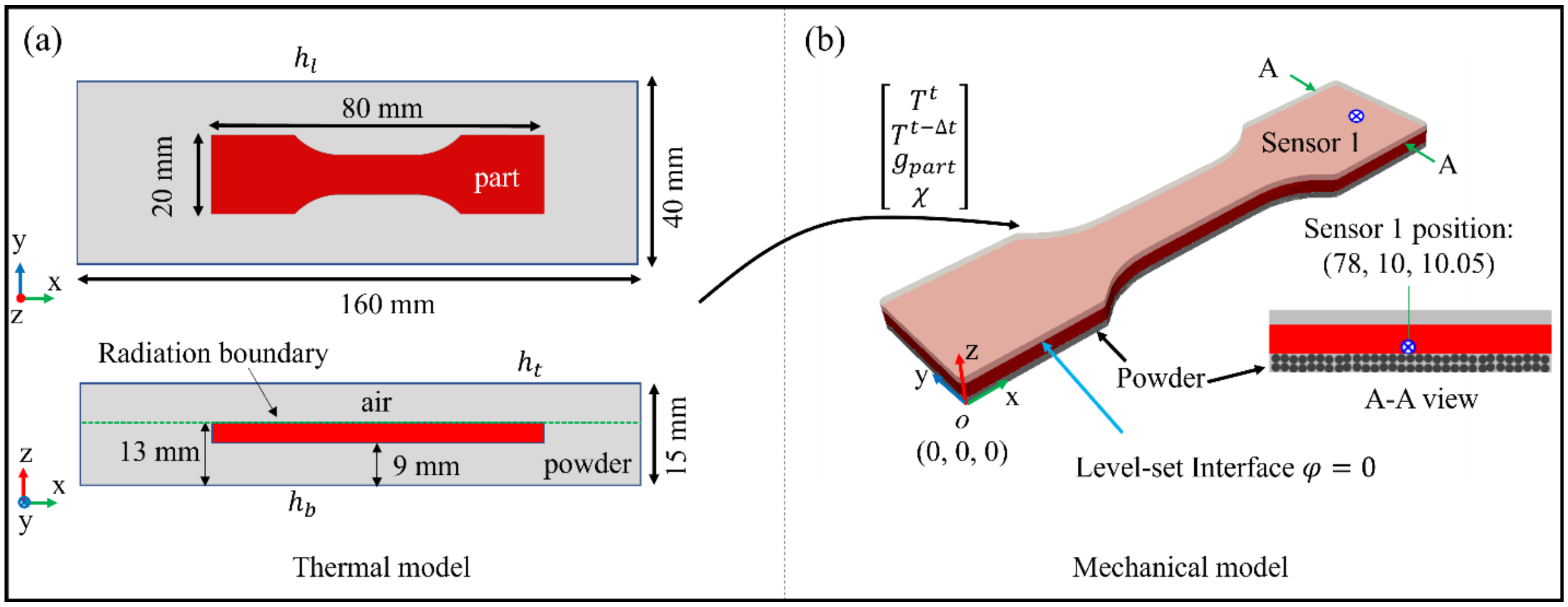

*Fig.2 Schematic configuration of the thermo-mechanical numerical model for the L-PBF process: (a) thermal model setup illustrating the sample location within the entire computational domain; (b) transfer of thermal fields to the mechanical model via a bi-mesh projection scheme.*

Fig.2 illustrates the configuration of the thermo-mechanical numerical model developed for simulating the L-PBF process of PA12. In Fig.2(a), the thermal model comprises multiple sub-domains including the printed part, surrounding non-melted powder, and ambient gas. To accurately capture the thermal interactions and interfaces between these regions, numerous fine elements are required, leading to a significantly increased number of elements in the whole mesh. While such complexity is computationally manageable for thermal simulations, it becomes prohibitively expensive for mechanical computations.

To address this, a simplified mesh configuration is adopted for mechanical analysis, as shown in Fig.2(b), in which the non-melted powder is omitted. This is justified by the limited mechanical contribution of the surrounding powder.

Thermal simulations are conducted using the thermal mesh (Fig.2(a)). At each time step, key thermal quantities namely the current and previous temperatures $T^t$and $T^{t-\Delta t}$, the phase fraction of the printed part $g_{part}$, and the relative degree of crystallinity $\chi$ are projected onto the mechanical mesh together (Fig.2(b)) for subsequent stress and deformation analysis.

To monitor localized field evolution, a virtual sensor (Sensor 1) is placed at the coordinate position (78, 10, 10.05), corresponding to the center of the first layer in the z-direction. In the xy-plane, it is located near the right edge of the sample rather than the center, as warpage phenomena are more prominently observed near the part boundaries during the L-PBF process. Please be noted that the single dog-bone numerical configuration in Fig.2 can be easily extended to other geometry such as with four dog-bone samples, as the counterpart for real experiment. Thermal boundary conditions are applied via effective heat transfer coefficients at the bottom ($h_b$), lateral ($h_l$), and top ($h_t$) surfaces.

Fig.3(a) presents the mesh employed for the thermal model. Coarse tetrahedral elements with an edge size of $h_{global}$ are assigned to the entire computational domain except for the top powder layer. The level-set method, together with its error estimation based mesh adaptation strategy [33], dynamically tracks the evolution of the powder layer during the recoating process. Fig.3(b) provides a closer view of the elements around the interface between the top powder layer and the surrounding air. The transition zone, with a thickness of $2\epsilon$, consists of approximately ten layers of fine elements with a mesh size of $h_{min}$. Fig.3(c) illustrates the mesh configuration used for the mechanical analysis, in which the global element size is identical to that of the thermal model. Two refined regions are defined to distinguish between (1) the bottom powder and the consolidated part, and (2) the material and the air. The corresponding element sizes and total element counts are summarized in Table 1. Fig.3(d) shows the locations of the four dog-bone samples; sensor 2 is placed in sample S2 to extract thermo-mechanical fields, and its relative position is identical to that of sensor 1 in the single-dog-bone configuration (see Fig.2).

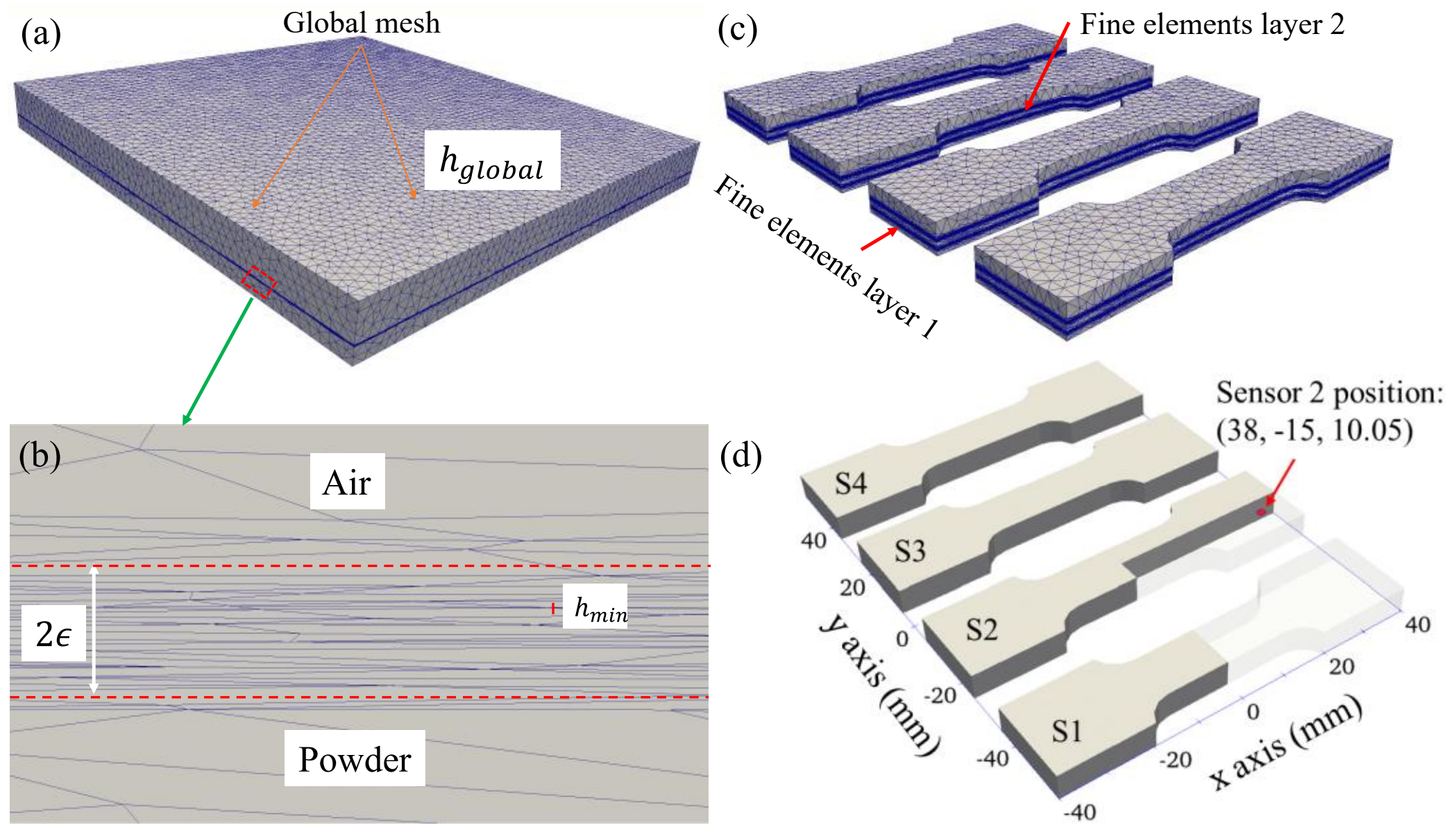

*Fig.3(a) The mesh for the thermal model; (b) closer view of the elements nearby the powder-air interface; (c) Mesh for the mechanical simulation; (d) Arrangement of four dog-bone samples and location of sensor 2 in sample S2, positioned analogously to sensor 1 in the single-dog-bone configuration (Fig.2).*

The convergence behavior of the thermal model was carefully verified at the part scale. The nonlinear Newton-Raphson iterative thermal solver demonstrates strong robustness owing to the stabilization scheme of the sub-incremental method proposed in our previous work [33]. To evaluate the convergence and mesh sensitivity of the mechanical simulation, numerical tests are conducted under a violent processing condition: $T_b = 155$ °C and $T_{new} = 130$ °C. Fig.4(a) shows the displacement (along the z-direction in the following context) recorded by Sensor 1 for different minimal element sizes $h_{min}$. It can be observed that the displacement evolution varies with $h_{min}$. Nevertheless, once the element size is reduced to 0.03 mm, the evolution curve tends to converge to a unique profile with only negligible differences. Furthermore, when $h_{min}$ = 0.03 mm is adopted, a highly accurate interpolation from the thermal mesh to the mechanical mesh is achieved, as illustrated in Fig.4(b). For each time step, the key physical fields including temperature and degree of crystallization exhibit excellent agreement between the two meshes.

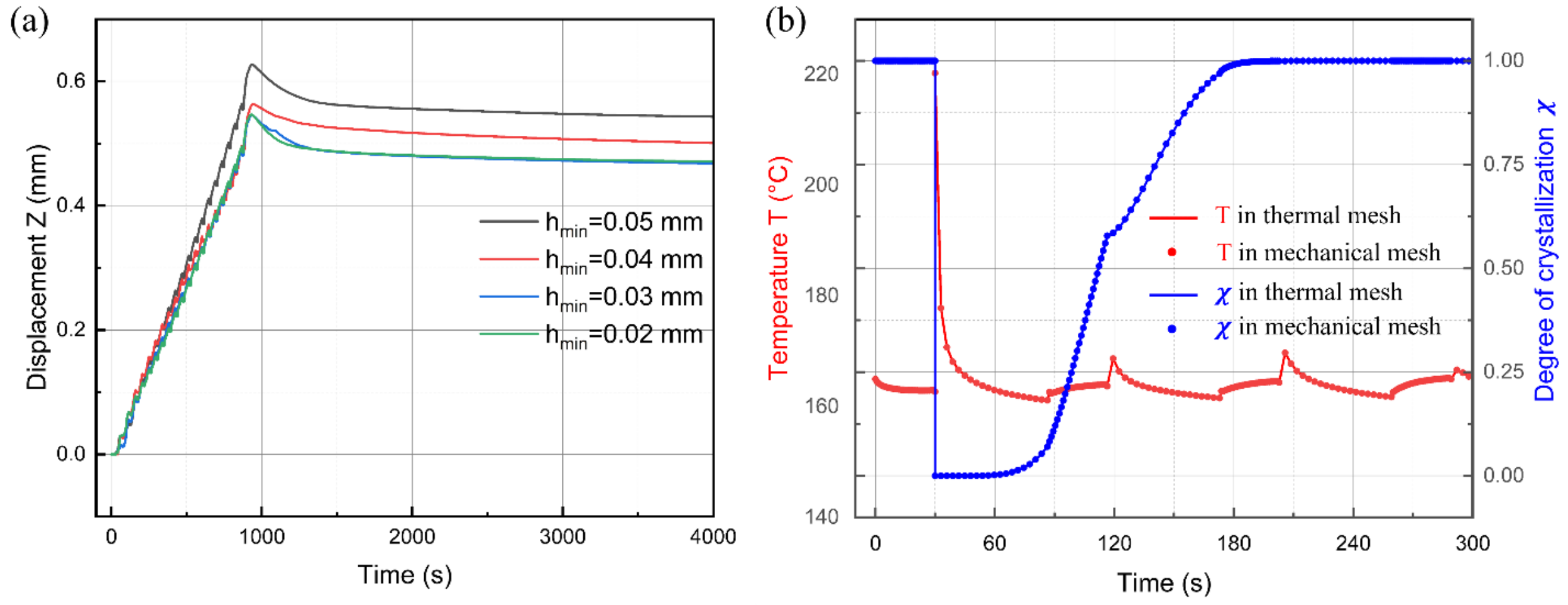

*Fig.4 (a) Mesh size sensitivity analysis of the mechanical simulation. (b) Evolution of key physical fields at the same position as Sensor 1 in the thermal and mechanical meshes.*

*Table 1 Mesh details of the thermal model and mechanical model*

| Mesh | $h_{min}$ [mm] | $h_{global}$ [mm] | Transition thickness $2\epsilon$ [mm] | Element number [million] | |
|---|---|---|---|---|---|
| | | | | with 1 dog-bone | with 4 dog-bones |
| Thermal model | 0.003 | 2.4 | 0.015 | 1.5 | 5.5 |
| Mechanical model | 0.02 | 2.4 | 0.2 | 0.3 | 1.0 |

Table 2 summarizes time step sizes for the various simulation stages. During the preheating stage, identical time increments were applied in all simulation cases. Because a relatively uniform and moderate heat flux was imposed to preheat the powder bed for 30 s, the time step was set to 1 s. In the heating stage, when the effective heat source model was employed, the total laser energy was deposited within a single time increment, corresponding to a time step of $6.25 \times 10^{-4}$ s. The cooling stage was resolved using 40 steps, whereas a 1 s increment was again used during the post-cooling phase. All simulations were executed in parallel on 64 CPU cores. At each time increment, the thermal model first solves the temperature field in a fully nonlinear manner. The resulting temperature field is subsequently used to evaluate additional physical quantities, including the degree of crystallization and the solidified fraction of the part. These computed fields, together with the temperature distribution, are then mapped to the mechanical model for stress and deformation analysis.

*Table 2 Summary of time step sizes used in different stages*

| Stage | Preheating | Scanning time $t_{scan}$ | | Post cooling |
|---|---|---|---|---|
| | | Heating $t_{heat}$ | Cooling $t_{cool}$ | |
| Time step size [s] | $\Delta t = 1$ | $\Delta t = \frac{\emptyset}{v} = 6.25\text{e} - 4$ | $\Delta t = \frac{t_{scan} - t_{heat}}{40} = 0.4$ | $\Delta t = 1$ |

## 2.2 Thermal modeling framework

The L-PBF process involves a series of complex coupled physical phenomena, among which the thermal history serves as the fundamental driving mechanism for phase transformations, material consolidation, and residual stress development. In numerical simulations, the transient temperature field is computed by solving the energy conservation equation expressed as:

$$\frac{\partial\{\rho h\}}{\partial t} = \nabla \cdot (\{\lambda\}\nabla T) + \dot{Q}_v + S_L - \dot{q} \tag{1}$$

where $\rho$ is the material density of PA12, $h$ is the temperature-dependent enthalpy incorporating sensible heat and $\lambda$ is the thermal conductivity. $S_L$ represents the latent heat associated with phase transformation, and $\dot{q}$ accounts for heat transfer between powder bed surface and the air in the printing champ if applicable.

The effective volumetric laser heat source $\dot{Q}_v$ that employed in macroscopic simulations, is developed based on the energy equivalence principle [31]:

$$P(1 - R)t_{scan} = \dot{Q}_v S t_{heat} \Delta Z \tag{2}$$

where $P$ is the nominal laser power, $R$ is the reflectivity, and $\Delta Z$ is the layer thickness. The heating time $t_{heat}$ represents the duration for which the moving laser beam scans over a material point. It is physically related to the laser beam diameter $\emptyset$ and the scanning velocity $v$ by [22,34,35]:

$$t_{heat} = \frac{\emptyset}{v} \tag{3}$$

while the scanning time per layer $t_{scan}$, is given by:

$$t_{scan} = \frac{S}{vH} \tag{4}$$

where $S$ is the scanned surface area of each layer, $H$ is the hatch spacing, more detailed explanations can be found in [33]. Substituting Eqs. (3, 4) into Eq. (2) yields:

$$\dot{Q}_v = \frac{P(1-R)}{\emptyset H \Delta Z} \tag{5}$$

The rest part of the $t_{scan}$ is considered as the cooling time: $t_{cool} = t_{scan} - t_{heat}$.

The heat loss through bottom, top and four lateral boundaries is considered with the convective heat exchange coefficient:

$$-\{\lambda\}\nabla \mathrm{T} \cdot n = h_c(T - T_{ext}) \tag{6}$$

where $n$ is the normal vector pointing outward. $T_{ext}$ is the surrounding temperature (25 °C).

**2.3 Mechanical model**

In this work, the mechanical response of the material is modeled using the force balance and mass conservation equations, which are solved within a velocity–pressure framework. The governing equations are expressed as:

$$\begin{cases} \nabla \cdot \boldsymbol{\sigma} + \boldsymbol{f_v} = 0 \\ \nabla \cdot \boldsymbol{u} - tr(\dot{\boldsymbol{\epsilon}}) = 0 \end{cases} \tag{7}$$

where $\boldsymbol{\sigma}$ is the Cauchy stress tensor, $\boldsymbol{f_v} = \rho \boldsymbol{g}$ denotes the gravity induced volume force. $\boldsymbol{u}$ is the velocity, $\dot{\boldsymbol{\epsilon}}$ represents the strain rate tensor. To solve the mechanical equilibrium equations, a constitutive relationship describing the material behavior is required and will be detailed in the following section.

Although plastic deformation may occur during L-PBF, it is neglected in this study due to the relatively low internal stresses arising from the absence of a rigid substrate constraint. As a result, the total strain rate is decomposed into three additive components:

$$\dot{\boldsymbol{\epsilon}} = \dot{\boldsymbol{\epsilon}}^{ve} + \dot{\boldsymbol{\epsilon}}^{v} + \dot{\boldsymbol{\epsilon}}^{cryst} + \dot{\boldsymbol{\epsilon}}^{th} \tag{8}$$

where $\dot{\boldsymbol{\epsilon}}^{\boldsymbol{ve}}$, $\dot{\boldsymbol{\epsilon}}^{v}$ are the viscoelastic and viscous strain rates, respectively. It should be noted that these two strain-rate components are mutually exclusive in this simulation. Specifically, in the fluid or powder state, only the viscous contribution $\dot{\boldsymbol{\epsilon}}^{v}$ is considered. The viscoelastic component $\dot{\boldsymbol{\epsilon}}^{ve}$ is activated once the molten polymer begins to solidify into a semicrystalline structure, as will be further linked to the degree of crystallization in the subsequent context. $\dot{\boldsymbol{\epsilon}}^{cryst}$is the crystallization-induced strain rate, $\dot{\boldsymbol{\epsilon}}^{th}$ is the thermal expansion strain rate. The thermal strain rate is defined based on isotropic assumption of the thermal expansion:

$$\dot{\boldsymbol{\epsilon}}^{th} = \mathrm{C}_{te}\dot{T}\mathbf{I} \tag{9}$$

where $\mathrm{C}_{te}$ is the linear coefficient of thermal expansion, and $\dot{T}$ is the time derivative of temperature, $\mathbf{I}$ is the second-order identity tensor.

The crystallization induced strain rate $\dot{\boldsymbol{\epsilon}}^{cryst}$ will be detailed in the following subsection.

For further analysis, the total strain rate tensor can be decomposed into deviatoric and volumetric components:

$$\begin{cases} \dot{\mathbf{e}} = \dot{\mathbf{e}}^{ve} + \dot{\mathbf{e}}^{v} \\ tr(\dot{\boldsymbol{\epsilon}}) = tr(\dot{\boldsymbol{\epsilon}}^{ve}) + tr(\dot{\boldsymbol{\epsilon}}^{v}) + tr(\dot{\boldsymbol{\epsilon}}^{cryst}) + tr(\dot{\boldsymbol{\epsilon}}^{th}) \end{cases} \tag{10}$$

here, $\dot{\mathbf{e}}$ is the deviatoric strain rate, which is assumed to arise from either viscous in apparent fluid state or viscoelastic effects in semicrystalline state. The volumetric strain rate includes contributions from viscoelastic deformation, viscous motion, crystallization-induced shrinkage, and thermal expansion.

## 2.4 Coupling with the modified crystallization kinetics

Phase transformation is one of the most critical phenomena in the AM of semi-crystalline polymers, primarily due to its dual physical consequences. First, crystallization or melting processes are typically accompanied by significant latent heat release or absorption, which strongly influences the thermal field. Second, due to the molecular rearrangement during crystallization, the material undergoes substantial specific volume changes, which may introduce internal stresses and ultimately compromise the dimensional accuracy of printed parts.

In recent years, crystallization kinetics models have been increasingly incorporated into numerical simulations of L-PBF for semi-crystalline polymers. Depending on the modeling objective and implementation, two types of crystallization models are commonly used: (i) the classic non-isothermal form, which is appropriate for the post-deposition cooling phase, and (ii) the modified kinetics models, specifically developed to account for the cyclic thermal history inherent to L-PBF processes. Both approaches are commonly based on the non-isothermal Nakamura model [28, 29].

In this work, we adopt a modified crystallization kinetics formulation proposed by Soldner et al. [22] and further calibrated in our previous study [30]. This model allows us to compute the transformed crystalline volume fraction $\chi$ as follows:

$$\frac{\partial \chi}{\partial t} = K(T) \cdot n(1-\chi)(-\ln(1-\chi))^{1-\frac{1}{n}} - \frac{\chi}{\Delta t}\left[1.0 - \left[0.5\left[\tanh\left(b(T_m^* - T)\right) + 1.0\right]\right]\right] \tag{11}$$

Here, $K(T)$ is a temperature-dependent rate constant, $n$ is the Avrami index, and $T_m^*$ represents the effective melting temperature threshold for suppressing crystallization during heating.

The crystallization-induced latent heat source term, which couples the thermal and phase evolution, is described by:

$$S_L = \rho \dot{\chi} \Delta h_c \tag{12}$$

where $\Delta h_c$ is the latent heat of crystallization, measured using differential scanning calorimetry (DSC).

Similarly, assuming that the specific volume change during phase transformation is linearly proportional to the transformed crystalline volume fraction [26], the crystallization-induced strain rate can be expressed as:

$$\dot{\boldsymbol{\epsilon}}^{cryst} = -\dot{\chi}\Delta V\mathbf{I} \tag{13}$$

where $\dot{\chi}$ is the rate of crystallization, $\Delta V$ denotes the total specific volume change associated with the solid-to-liquid (or vice versa) phase transition. **I** is the identity tensor, indicating that the strain is assumed to be isotropic. This formulation establishes a direct link between the evolution of the solid fraction and the volumetric strain induced by crystallization, and is particularly relevant for semi-crystalline polymers undergoing rapid thermal cycles during laser-based AM processes.

### 2.5 Constitutive law and material properties

The mechanical modeling of the L-PBF process necessitates a multi-domain, multi-phase constitutive framework. The overall simulation system comprises three distinct domains: the surrounding air, the laser-processed region, and the underlying powder bed. The polymer initially exists in a powder state prior to melting. Upon laser irradiation, the processed region undergoes a transition from the molten to the semi-crystalline phase, with the phase evolution governed by crystallization kinetics.

#### 2.5.1 Constitutive formulation

In this study, both the underlying powder bed and the polymer in its molten state (defined by $\chi < 0.1$) are treated as viscous fluids:

$$\mathbf{S} = 2\mu\left(\dot{\boldsymbol{\epsilon}} - \frac{1}{3}tr(\dot{\boldsymbol{\epsilon}})\mathbf{I}\right) \tag{14}$$

where $\mu$ is the dynamic viscosity, the stress deviator $\mathbf{S} = \boldsymbol{\sigma} + p\mathbf{I}$ (with pressure $p = -\frac{1}{3}tr(\boldsymbol{\sigma})$). The crystallized region is treated as a solid exhibiting viscoelastic behavior over a broad temperature range.

The time-dependent stress response of semi-crystalline polymers during L-PBF is characterized by a linear viscoelastic framework. The constitutive relation in the time domain can be represented by linking the Gauchy stress tensor $\boldsymbol{\sigma}(t)$ to the history of the viscoelastic strain $\boldsymbol{\epsilon}^{ve}$ through the integral equation [38]:

$$\boldsymbol{\sigma}(t) = \int_{-\infty}^{t} \mathbb{E}(t-\tau):\frac{\partial \boldsymbol{\epsilon}^{ve}}{\partial \tau}d\tau \tag{15}$$

where $\mathbb{E}(t)$ is the fourth-order relaxation stiffness tensor, which can be decomposed into deviatoric and volumetric components as:

$$\mathbb{E} = 2G(t)\mathbb{I}^{dev} + 3K(t)\mathbb{I}^{vol} \tag{16}$$

The deviatoric $\mathbb{I}^{dev}$ and volumetric $\mathbb{I}^{vol}$ projection operators are defined in terms of the second-order identity tensor **I** as**:**

$$\begin{cases} \mathbb{I}^{vol} = \dfrac{1}{3}\mathbf{I}\otimes\mathbf{I} \\ \mathbb{I}^{dev} = \mathbb{I} - \mathbb{I}^{vol} \end{cases} \tag{17}$$

$G(t)$ and $K(t)$ denote the time-dependent shear and bulk moduli, respectively. Their temporal evolution is modeled using the Generalized Maxwell Model (GMM) [39], expressed through Prony series as:

$$\begin{cases} G(t) = G_\infty + \sum_{i=1}^{I} G_i \, exp\left(-\frac{t}{g_i}\right) \\ K(t) = K_\infty + \sum_{j=1}^{J} K_j \, exp\left(-\frac{t}{k_j}\right) \end{cases} \tag{18}$$

In these equations, $I, J$ are the number of Maxwell branches for shear and volumetric response, respectively. $G_\infty$ and $K_\infty$ are the long-term (equilibrium) shear and bulk moduli, while $G_i$, $K_j$, $g_i$, and $k_j$ denote the modulus and relaxation time of each branch. Notably, the numbers of branches for shear and bulk may differ.

Meanwhile, under the assumption of constant strain rate over a discrete time increment $\Delta t = t_{n+1} - \boldsymbol{t_n}$, the hereditary integral (Eq. 15) can be evaluated in a recursive form, yielding the deviatoric and hydrostatic stress updates:

$$\begin{cases} \mathbf{s}(t_{n+1}) = 2G_\infty \mathbf{e}^{\boldsymbol{ve}}(t_{\boldsymbol{n}}) + 2\sum_{i=1}^{I} G_i \left[1 - \exp\left(-\frac{\Delta t}{g_i}\right)\right] \frac{g_i}{\Delta t} \Delta \mathbf{e}^{\boldsymbol{ve}} + \sum_{i=1}^{I} \exp\left(-\frac{\Delta t}{g_i}\right) \mathbf{s_i}(t_{\boldsymbol{n}}) \\ \sigma_H(t_{\boldsymbol{n+1}}) = 3K_\infty \epsilon_H^{ve}(t_{\boldsymbol{n}}) + 3\sum_{j=1}^{J} K_j \left[1 - \exp\left(-\frac{\Delta t}{k_j}\right)\right] \frac{k_j}{\Delta t} \Delta \epsilon_H^{ve} + \sum_{j=1}^{J} \exp\left(-\frac{\Delta t}{k_j}\right) \sigma_{H\,j}(t_{\boldsymbol{n}}) \end{cases} \tag{19}$$

where $\mathbf{s}(t)$ is the deviatoric part of the stress tensor, $\sigma_H(t_{\boldsymbol{n}})$ is the hydrostatic (mean normal) stress, which corresponds to the pressure $p$. $\mathbf{e}^{\boldsymbol{ve}}$, $\epsilon_H^{ve}$ are the deviatoric and volumetric parts of the viscoelastic strain.

In practice, accurate representation of polymer rheology requires many branches in the Prony series to fit stress relaxation over a wide timescale. However, determining these parameters over a range of temperatures and strain rates can be challenging. To address this, the time-temperature superposition principle is introduced. By defining a reduced time $\xi$ [40,41], the influence of temperature on viscoelastic behavior is captured through a shift factor $a_T$, such that:

$$\Delta\xi = \frac{\Delta t}{a_T(T(t), T_{\text{ref}})} \tag{20}$$

where $T(t)$ is the current temperature and $T_{\text{ref}}$ is a reference temperature. The shift factor $a_T$ accounts for the thermo-rheological behavior of the polymer and can be obtained from experimental data. Therefore, at any arbitrary temperature conditions, the discretization form of the deviatoric and spherical stresses can be re-organized by substituting Eq. 20 into Eq. 19:

$$\begin{cases} \mathbf{s}(t_{n+1}) = 2G_\infty \mathbf{e}^{ve}(t_n) + 2\sum_{i=1}^{I} G_i \left[1 - \exp\left(-\frac{\Delta\xi \cdot a_T}{g_i}\right)\right] \frac{g_i}{\Delta\xi \cdot a_T} \Delta\mathbf{e}^{ve} + \sum_{i=1}^{I} \exp\left(-\frac{\Delta\xi \cdot a_T}{g_i}\right) \mathbf{s}_i(t_n) \\ \sigma_H(t_{n+1}) = 3K_\infty \epsilon_H^{ve}(t_n) + 3\sum_{j=1}^{J} K_j \left[1 - \exp\left(-\frac{\Delta\xi \cdot a_T}{k_j}\right)\right] \frac{k_j}{\Delta\xi \cdot a_T} \Delta\epsilon_H^{ve} + \sum_{j=1}^{J} \exp\left(-\frac{\Delta\xi \cdot a_T}{k_j}\right) \sigma_{Hj}(t_n) \end{cases} \tag{21}$$

This formulation enables the unified description of thermally dependent viscoelastic response under varying temperature histories, which is critical for simulating realistic thermo-mechanical interactions in the L-PBF process.

### 2.5.2 Thermo-viscoelasticity characterization

The viscoelastic behavior of the material was characterized using a dynamic mechanical analyzer (DMA, Bohlin Tritec 2000) operated in tensile mode. Rectangular specimens of L-PBF-printed PA12 with dimensions of 12 × 5 × 2 $mm^3$ were subjected to a sinusoidal oscillatory strain to determine their storage and loss moduli.

The DMA specimens were fabricated using a Sintratec Kit L-PBF system (Sintratec Kit, AG, Brugg, Switzerland). The processing parameters were as follows: preheating target temperature of 165 °C, powder tank temperature of 140 °C, laser power of 2.3 W, scanning velocity of 400 mm/s, and hatch spacing of 200 μm, as summarized in Table 4. Four rectangular samples were built at the center of the build platform to minimize spatial variations in the thermal field. A cross-hatching scanning strategy with a 90° rotation between consecutive layers was employed, which is the default pattern of the Sintratec system.

To ensure that the material response remained within the linear viscoelastic regime, the displacement amplitude was fixed at 0.005 mm, resulting in a maximum strain below $10^{-3}$.

A frequency sweep experiment was performed across a temperature range from 30 °C to 170 °C with 5 °C increments. At each isothermal condition, the sample was tested over a frequency range of 0.1–10 Hz. To eliminate potential compressive effects during testing, a tensile preload of 1 N was applied to each specimen. A total of three specimens were subjected to DMA testing to ensure measurement repeatability. The averaged storage and loss moduli obtained from these tests were used to construct the corresponding master curves.

Fig.5(a) displays the averaged storage modulus data across various isothermal temperatures. A pronounced modulus drop is observed in the range of 40–60 °C, suggesting the occurrence of a glass transition. Based on this observation, 55 °C was selected as the reference temperature ($T_{\mathrm{ref}}$) for constructing the master curve.

Using the time-temperature superposition (TTS) principle [42], storage modulus at temperatures adjacent to $T_{\mathrm{ref}}$ were horizontally shifted by corresponding shift factors ($a_T$) to align with the reference curve. As shown in Fig.5(a), a smooth master curve was successfully constructed, capturing the mechanical response over an extended frequency domain. The corresponding logarithmic shift factors, $\log(a_T)$, were plotted as a function of temperature in Fig.5 (b).

Unlike classical polymers that exhibit monotonic Williams–Landel–Ferry (WLF) [43] or Arrhenius [44] behavior, the $\log(a_T)$ profile in this case shows different evolution trends—initially concave and then convex—which deviates from both WLF and Arrhenius model predictions.

Consequently, a direct linear interpolation approach was adopted to estimate $a_T$ values between adjacent temperature points.

The resulting master curve in Fig.5(a) provides a comprehensive description of the frequency-dependent mechanical response. Following the methodology proposed in [35, 36], this response was further fitted using GMM with a series of Prony terms.

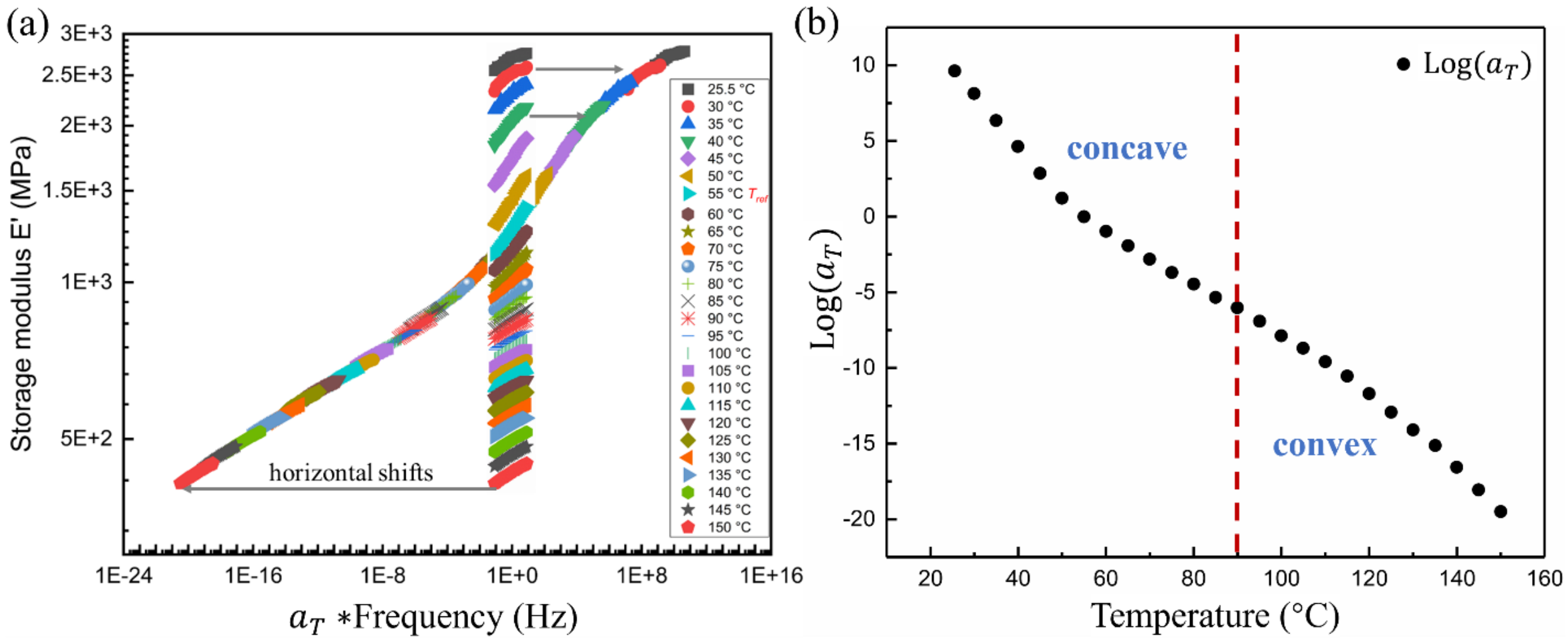

*Fig.5 (a) Master curve of storage modulus E' for PBF-PA12 across 25.5–150 °C, constructed with reference temperature is 55°C. (b) The corresponding logarithmic shift factor as a function of temperature.*

Fig.6(a) illustrates the frequency-dependent evolution of the storage modulus ($E'$) and loss modulus ($E''$). The red curves represent the experimental data obtained from DMA, while the colored markers denote the predictions from the GMM with different numbers of Prony branches (N). When N=10, the model exhibits noticeable oscillations and substantial deviations from the experimental results, indicating insufficient fitting accuracy. Increasing N to 20 and 40 markedly improves the fitting performance, capturing the experimental trends with satisfactory agreement, although minor discrepancies remain in both moduli.

Fig.6(b) presents the relative errors between the experimental measurements and the GMM predictions. Large errors are observed across the entire frequency range when N=10, confirming the inadequate representation of the viscoelastic behavior. When N=20 and 40, the relative errors first increase with frequency and then decrease, following similar patterns. The model with 40 branches provides marginal improvement in the high-frequency regime, suggesting that further increasing N yields diminishing returns.

Additional processing simulations with N=20 and N=40 show nearly identical displacement predictions. Nevertheless, in the GMM implementation, each branch requires storage of a nine-component stress tensor. With no noticeable improvement in the numerical results, the simulation with N=40 required approximately 121% of the computational time compared to that with N=20. Therefore, considering both computational efficiency and accuracy, N=20 was adopted in the subsequent simulations.

The fitted Prony series parameters, including the moduli $E_i$ and corresponding relaxation times $\tau_i$, are summarized in Table 3. It should be noted that the tensile mode DMA derived modulus values $E_i$ are first converted into shear moduli $G_i$ and bulk moduli $K_i$ before being implemented into the mechanical solver. This conversion is performed using the Poisson's ratio as follows:

$$G_i = \frac{E_i}{2(1+\nu)}, K_i = \frac{E_i}{3(1-2\nu)} \tag{22}$$

In thermoplastic polymers, Poisson's ratio is known to be a time-dependent parameter and, strictly speaking, is not a true material constant. However, due to the lack of sufficient experimental data, a constant Poisson's ratio of 0.3 [39,47] was assumed in this study. This simplification allows the shear and bulk relaxation moduli to be derived from the tensile relaxation modulus according to the classical isotropic relations.

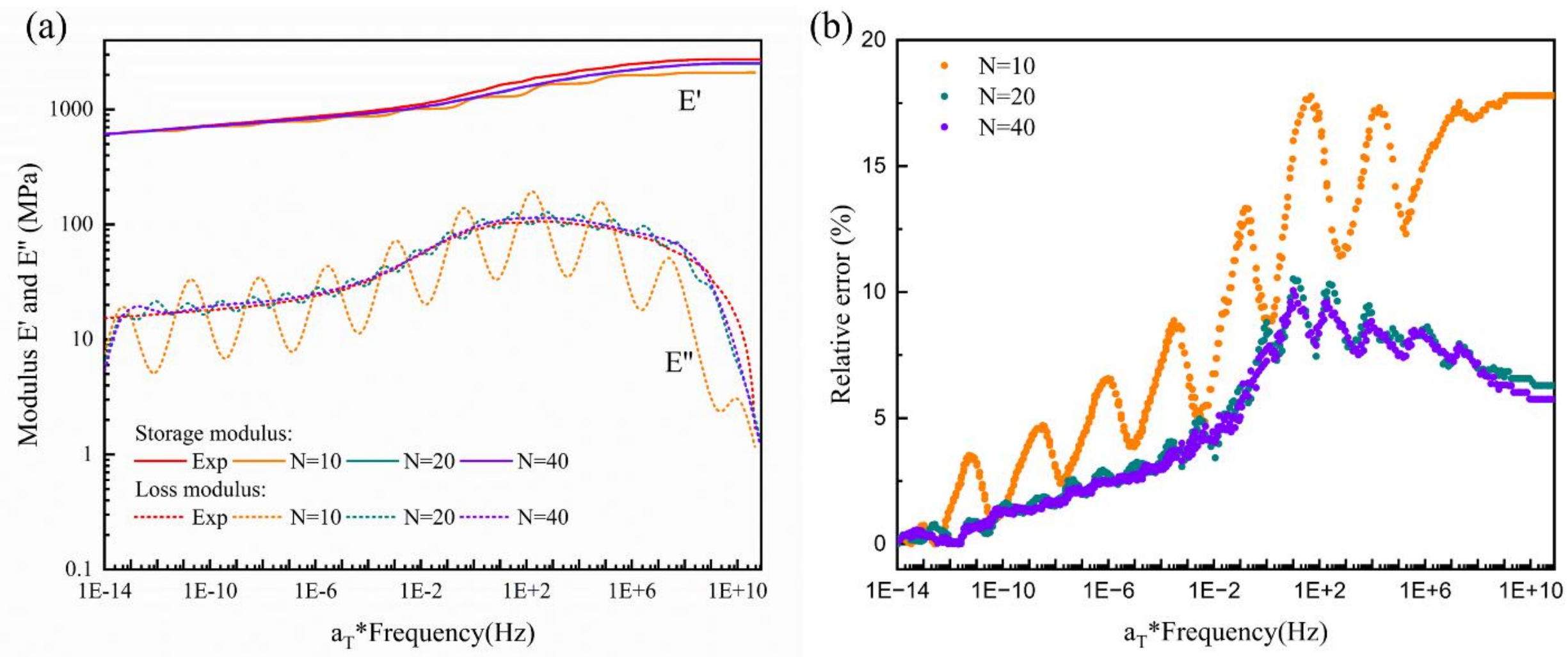

*Fig.6 (a) Frequency-dependent evolution of storage modulus (E') and the loss modulus (E''). "Exp" refers to DMA test data; "N=10, 20, 40" denote the GMM with N branches. (b) Relative errors of the storage modulus between the GMM predictions and the experimental results.*

*Table 3 Identified Prony parameters of the GMM with 20 branches for PBF-PA12 viscoelasticity*

| Branch i | 1 | 2 | 3 | 4 | 5 | 6 | 7 | 8 | 9 | 10 |
|---|---|---|---|---|---|---|---|---|---|---|
| $E_i$ [MPa] | 59 | 67 | 109 | 148 | 173 | 185 | 186 | 180 | 160 | 129 |
| log ($\tau_i$/s) | -11.7 | -9.8 | -8.25 | -6.76 | -5.34 | -4.0 | -2.75 | -1.57 | -0.46 | 0.65 |
| i | 11 | 12 | 13 | 14 | 15 | 16 | 17 | 18 | 19 | 20 |
| $E_i$ [MPa] | 99 | 72 | 55 | 43 | 35 | 31 | 29 | 26 | 5 | 181 |
| log ($\tau_i$/s) | 1.77 | 2.93 | 4.11 | 5.30 | 6.48 | 7.63 | 8.69 | 9.56 | 10.47 | 11.58 |

### 2.5.2 Material properties

The commercially available black PA12 powder (Sintratec) [48] is employed in this study for both material characterization and L-PBF process simulation. The material properties used in the numerical model are summarized in Table 4.

For the polymer phase, the latent heat of crystallization was measured to be 60500 $\mathrm{J \cdot kg^{-1}}$[30], The primary melting range and remelting temperature of PA12 were determined as 184-192 °C and 178 °C, respectively, using DSC. The temperature-dependent dynamic viscosity of molten PA12 was characterized using a parallel-plate shear rheometer. Since the densification process is not considered in the present simulation, a constant density of 1000 $\mathrm{kg \cdot m^{-3}}$ is assumed for the polymer. The dynamic viscosity of air is assumed to be $2.0 \times 10^{-5}$ Pa·s [49], while the temperature-dependent viscosity of PA12 was experimentally measured, as detailed in [30].

Regarding thermal dilatational behavior, a constant thermal expansion coefficient of $1.2 \times 10^{-4}$ $\mathrm{K^{-1}}$ is used, along with a crystallization-induced specific volume change of $\Delta V = 0.047\ \mathrm{cm^3/g}$ as characterized in [15]. The temperature-dependent thermal conductivity of PA12 is adopted from the experimental results reported by Riedlbauer et al. [28], where a distinct increase in conductivity was observed during the melting transition.

In addition, the thermophysical properties of air and the laser processing conditions are also listed in Table 4.

*Table 4 Material properties and common processing parameters used in this work*

| Object | Property | Symbol | Value | Unit | Source |
|---|---|---|---|---|---|
| PA12 | Crystallization latent heat | $\Delta h_c$ | 60500 | $\mathrm{J \cdot kg^{-1}}$ | Exp[30] |
| | Melting range | $T_{im} - T_{fm}$ | 184-192 | °C | |
| | Remelting temperature | $T_m^*$ | 178 | °C | |
| | Viscosity | $\eta_m$ | Fig.1(d) in ref | $\mathrm{Pa \cdot s}$ | |
| | Specific heat | $C_{p,m}$ | 2650 | $\mathrm{J \cdot kg^{-1} \cdot K^{-1}}$ | [25] |
| | Density PA12 | $\rho_m$ | 1000 | $\mathrm{kg \cdot m^{-3}}$ | [50] |
| | Thermal conductivity | $\lambda_m$ | Fig.5 in ref | $\mathrm{W \cdot m^{-1} \cdot K^{-1}}$ | [28] |
| | Specific volume change | $\Delta V$ | 0.047 | $\mathrm{cm^3 \cdot g^{-1}}$ | [15] |
| | Thermal expansion coefficient | $C_{te}$ | $1.2 \times 10^{-4}$ | $\mathrm{K^{-1}}$ | [51] |
| Air | Heat capacity | $C_{p,a}$ | 1006 | $\mathrm{J \cdot kg^{-1} \cdot K^{-1}}$ | [52] |
| | Density | $\rho_a$ | 1.208 | $\mathrm{kg \cdot m^{-3}}$ | |
| | Thermal conductivity | $\lambda_a$ | 0.024 | $\mathrm{W \cdot m^{-1} \cdot K^{-1}}$ | |
| | Viscosity | $\eta_a$ | $2.0 \times 10^{-5}$ | $\mathrm{Pa \cdot s}$ | [49] |
| Processing parameters | Laser power | $P$ | 2.3 | W | [53] |
| | Laser diameter | $\emptyset$ | 250 | μm | |
| | Reflection coefficient | $R$ | 0.1 | | [30] |
| | Scan velocity | $V$ | 400 | mm/s | User defined |
| | Hatch distance | $H$ | 0.2 | mm | |
| | Layer thickness | $\Delta Z$ | 0.1 | mm | |
| | Thermal exchange coefficient | $h_b/h_l/h_t$ | 10/5/5 | $\mathrm{W \cdot m^{-2} \cdot K^{-1}}$ | [33] |

## 3. Results and discussion

The developed thermo-mechanical numerical model is employed to investigate several critical physical phenomena governing warpage in the L-PBF of semi-crystalline polymers. Fig.7 provides a schematic representation of the L-PBF process, including the key thermo-mechanical interactions and the research objectives addressed in this study.

First, despite the recognized importance of the powder bed in supporting the part during fabrication, the mechanical behavior of the pre-deposited powder beneath the processed region, indicated by the red outline in Fig.7, has been largely overlooked in previous modeling efforts. Many existing studies simply impose a fixed boundary condition on this region, which may lead to ambiguous or physically inconsistent simulation results. Sub-section 3.1 explores three alternative boundary conditions to more accurately capture the mechanical response of the support powder.

Second, although both thermal expansion and crystallization-induced volume contraction can contribute to part displacement and warpage, their respective roles and interactions remain insufficiently quantified in the literature. Moreover, the stress evolution associated with phase transformation—especially during the post-cooling stage—has not been thoroughly characterized. These issues are systematically analyzed in Sections 3.2 and 3.3.

Third, as illustrated in Fig.7, the preheating process in polymer L-PBF involves two independent thermal control parameters: the preheating target temperature ($T_b$) and the temperature of the recoated powder ($T_{new}$). Although elevated preheating conditions can reduce uneven shrinkage and warpage, they also pose the risk of premature material aging or degradation [54]. Section 3.4 aims to investigate the influence of both $T_b$ and $T_{new}$ on the thermal history and deformation behavior.

Lastly, radiative heat transfer between the powder bed surface and the surrounding build chamber can significantly influence crystallization kinetics and, consequently, part warpage during the LPBF process. However, its effect has not been explicitly considered or thoroughly justified in most existing numerical models [55,56]. In Section 3.5, a detailed investigation is conducted to assess the impact of radiative boundary conditions.

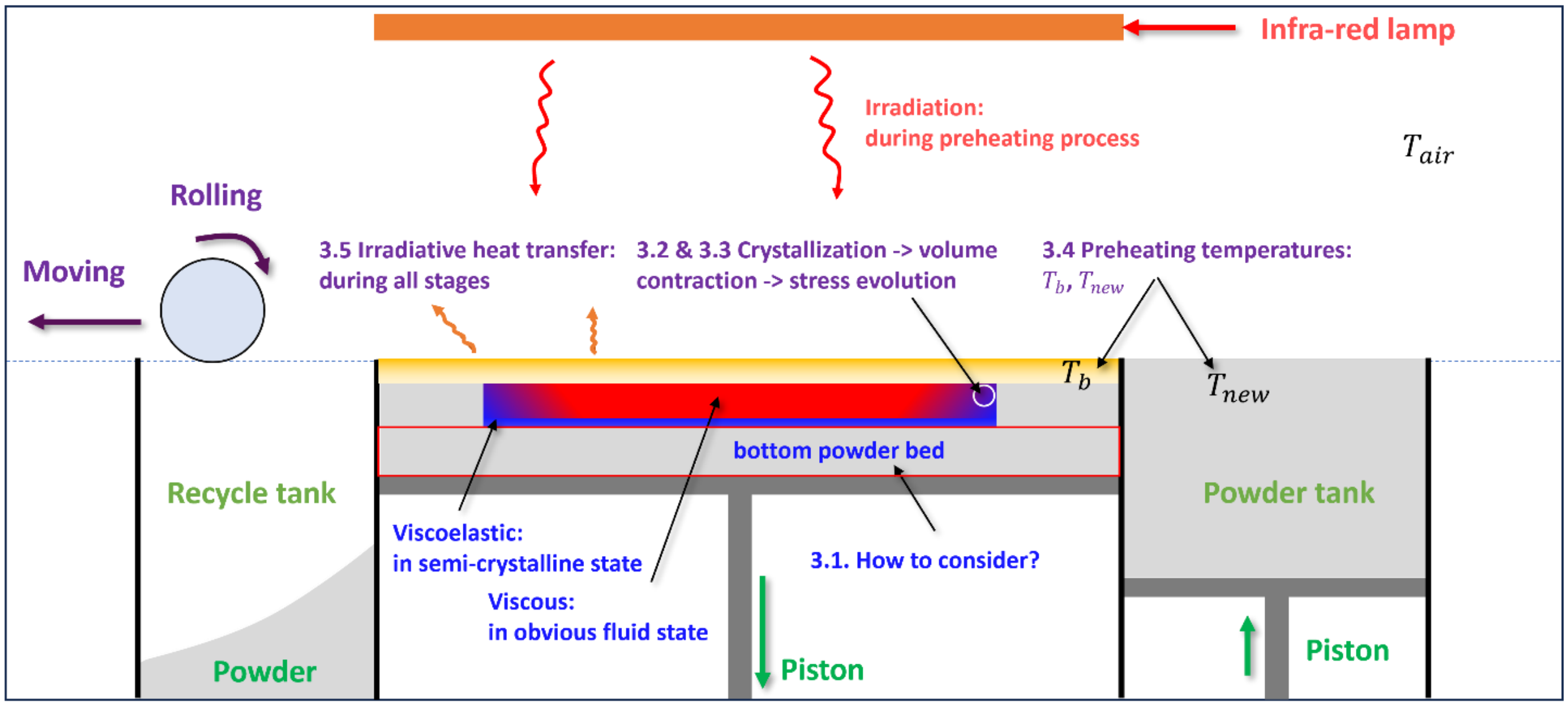

*Fig.7 Schematic illustration of key physical mechanisms and research objectives addressed in this study for the L-PBF process of semi-crystalline polymers.*

## 3.1. Mechanical boundary conditions

In this section, we examine three alternative mechanical boundary conditions to represent the support powder and analyze their effects on part displacement and final warpage.

Fig.8 (a–c) illustrates the three modeling strategies for the bottom powder bed located beneath the specimen, as indicated by the red outlines. Fig.8 (d–f) presents the corresponding displacement fields and resulting warpage. To ensure consistent thermal conditions in all simulations, the build platform and surrounding powder bed were initialized at a uniform preheating temperature of 160 °C, representative of typical polymer L-PBF settings. This value serves as the baseline initial thermal condition for evaluating the influence of different mechanical constraints in Fig.8.

In Fig.8 (a), all nodes representing the support powder are fixed. As a result, the bottom region exhibits negligible displacement, and the final part experiences almost no deformation. In Fig.8 (b), only the central zone beneath the consolidated region is fixed, while the surrounding powder is left unconstrained. This leads to upward displacement in the center and downward movement of the lateral edges. This deformation results from the layer-wise crystallization process typical in polymer L-PBF: the lower layers crystallize first, shrink earlier, and induce a downward bending of the part. However, this warping direction contradicts experimental observations, where upward warpage at the edges is more commonly observed.

In Fig.8 (c), a more realistic semi-supported condition is established by modeling the underlying powder bed as a fictitious 1 mm-thick viscous layer with a viscosity of 1 MPa·s, providing sufficient support for the molten polymer. This viscous medium suppresses early-stage deformation and counteracts gravity, thereby reducing downward sinking and warpage. During cooling stage, as the upper layers begin to crystallize and stiffen, the part bends upward. Although the viscous layer continues to resist this upward deformation, its constraining effect gradually weakens with the increasing stiffness of the solidified material. Overall, this configuration better captures the experimentally observed warpage pattern.

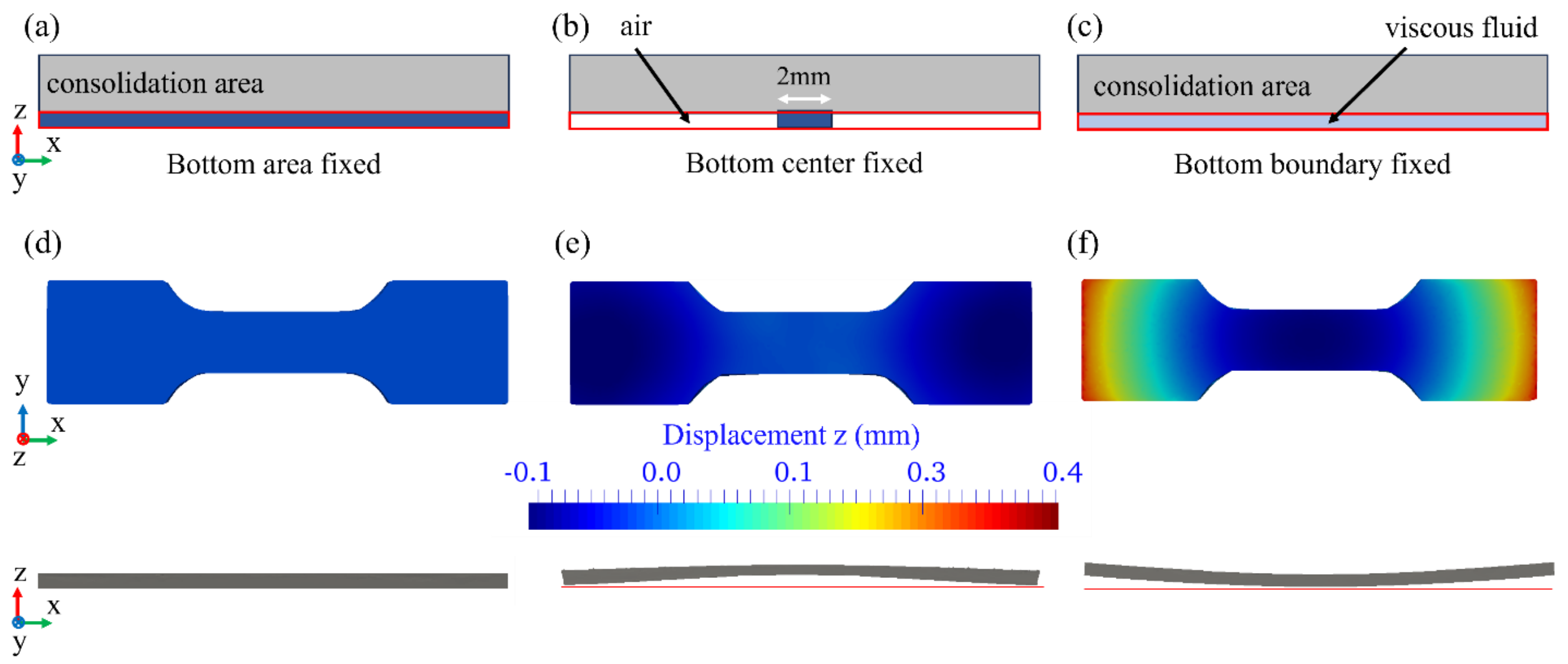

*Fig.8 Mechanical boundary condition strategies for modeling the bottom powder bed and their effects on part deformation. (a) Fully constrained support powder region (1 mm thick); (b) only the central zone beneath the part is fixed; (c) the support powder is modeled as a 1 mm-thick viscous layer (1 MPa·s). The red outlines indicate the powder bed region beneath the specimen. (d–f) Corresponding simulated displacement fields and resulting warpage (magnified by a factor of 5) for each boundary condition.*

### 3.2. Dominant Role of Phase Transformation Induced Volume Change.

This section analyzes the respective and combined effects of thermal expansion and crystallization induced specific volume change on the part's displacement.

Fig.9 presents the temperature and displacement evolutions at the right wing of the first deposited layer, a region frequently exhibiting curling behavior. The data are captured at Sensor 1 (as presented in Fig.2(b)), which is positioned to monitor this critical area. The upper plots in Fig.9 provide a zoomed-in view of the corresponding lower plots, highlighting the detailed variations during the layer-by-layer process.

Fig.9 (a) and (b) correspond to the phenomenon (Case1) where only thermal expansion strain rate ($\dot{\epsilon}^{th}$) is considered. In Case2, Fig.9 (c) and (d) depict results obtained by accounting solely for the crystallization-induced volume change ($\dot{\epsilon}^{cryst}$). In Fig.9(e) and Fig.9 (f), both thermal expansion and crystallization-induced shrinkage are incorporated simultaneously in Case3. Fig.9(a), (c), and (e) highlight the displacement evolution during the printing stage, while Fig.9(b), (d), and (f) show the displacement trends throughout the entire thermal history, including both the printing and subsequent cooling stages.

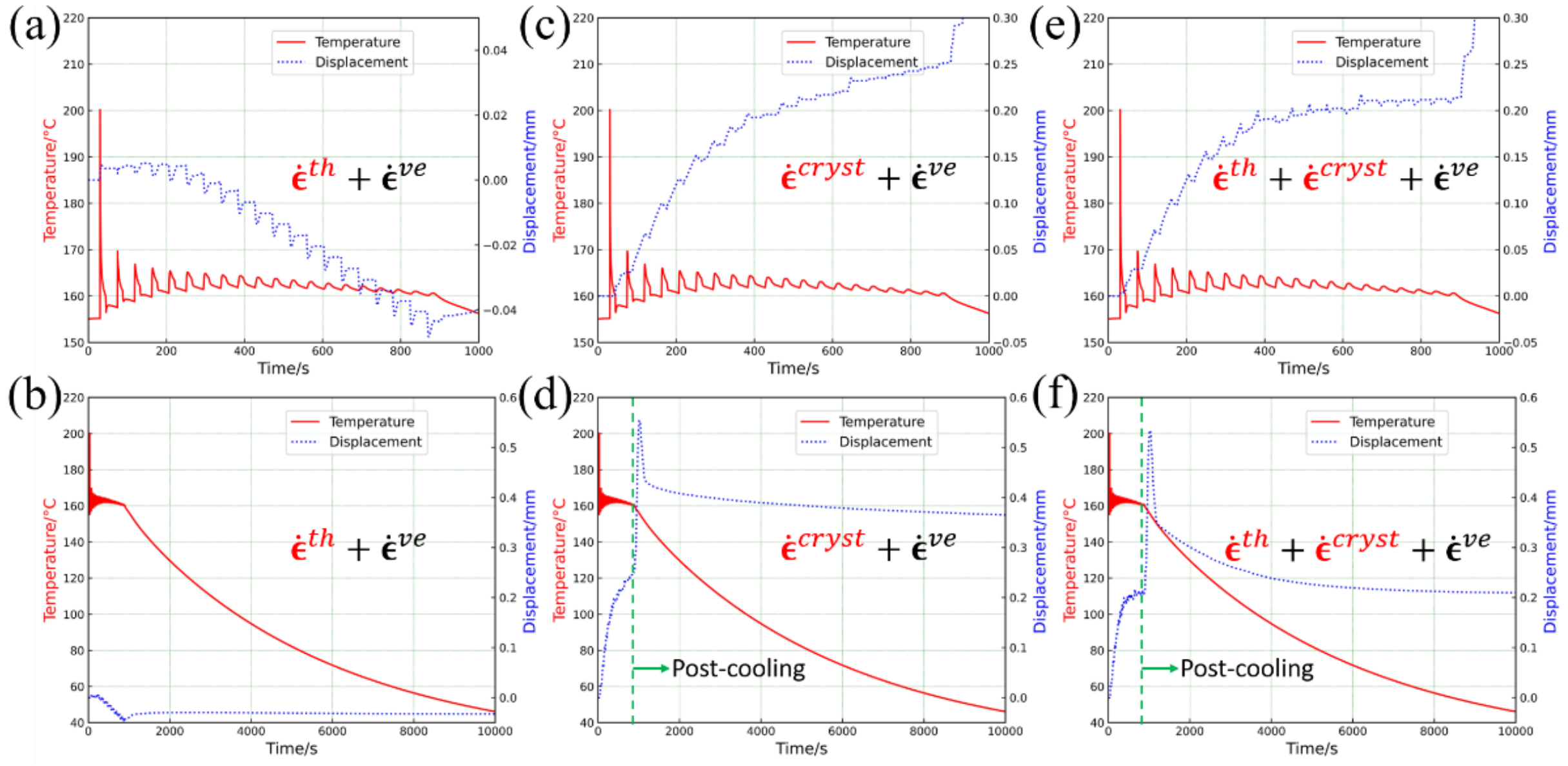

*Fig.9 Displacement and temperature evolution at Sensor 1 (right wing of the first layer), comparing three modeling scenarios:(a, b) Thermal expansion only;(c, d) Crystallization-induced volume change only; (e, f) Combined effects. First row (a, c, e): printing stage; Second row (b, d, f): entire process including cooling.*

To ensure consistency, all three cases in Fig.9 employ the same thermal model, including identical material properties and processing parameters. As a result, the temperature histories across the cases are identical. Nevertheless, despite the same thermal evolution, the mechanical responses vary significantly due to the different assumptions in strain decomposition, i.e., different combination of strain rates $\dot{\epsilon}^{th}$, $\dot{\epsilon}^{cryst}$ in Eq.8.

Fig.9(a) shows the displacement response of Case 1, where only thermal expansion is considered. Each energy deposition event leads to a temperature increase, inducing thermal expansion and thus a downward movement of the sensor (negative displacement). Subsequently, as the material cools, the laser processed region gradually contracts, causing a slight upward recovery. Additionally, the deposition of a new powder layer, which is colder than the current surface, introduces further volume contraction, leading to a net positive displacement. During the post-cooling phase, the displacement stabilizes after a minor rebound, as illustrated in Fig.9(b).

In Case 2, only the crystallization-induced volume change is accounted for. The displacement evolution in Fig.9(c) reveals that crystallization plays a dominant role in shaping the deformation of the semi-crystalline polymer. Under the same thermal conditions, the maximum displacement is more than an order of magnitude greater than that in Case 1, underscoring the critical importance of phase transformation. Unlike Case 1, the displacement during printing is not monotonic. It increases during both cooling and new-layer deposition, driven by accelerated crystallization. Conversely, energy input during laser scanning tends to slow down or partially remelt the crystalline zones, temporarily reducing the displacement.

Fig.9(e) presents the result of Case 3, which considers both $\dot{\epsilon}^{th}$ and $\dot{\epsilon}^{cryst}$. Compared to Case 2, the overall displacement amplitude is slightly reduced, likely due to the partial compensation between thermal expansion and crystallization-induced shrinkage. However, significant deformation still accumulates throughout the printing process, indicating a persistent risk of layer failure or powder recoating defects in L-PBF.

In both Case 2 and Case 3, after the energy input ceases, the vertical displacement exhibits a sharp increase, reaching its peak at approximately 1022 s. This rise is likely caused by intense crystallization occurring near the edge region of the upper layers, which induces nonuniform volumetric shrinkage. The subsequent rapid decline in displacement corresponds to the phase transformation in the central region, as illustrated in Fig.10. Thereafter, the displacement continues to decrease gradually even after crystallization has completed (at about 1400 s). Since crystallization is no longer active, this post-transformation deformation can be attributed to internal stress relaxation within the solidified polymer. Notably, in Case 2 (Fig.9(d)), thermal expansion is explicitly excluded from the strain formulation, further confirming that residual-stress relaxation is the primary driver of the late-stage displacement recovery. The detailed analysis of the stress evolution is presented in the following section.

### 3.3. Stress State Analysis

To elucidate the underlying physical mechanisms, Fig.10 presents the temporal evolution of the relative degree of crystallization, crystallization rate, and the spatial distribution of the normal stress component $\sigma_{xx}$ (top view) for Case 2, where only crystallization-induced volume change ($\dot{\epsilon}^{cryst}$) is considered.

At $t = 882$ s, corresponding to the completion of the final layer deposition, solidification is observed only near the shoulders of the dog-bone specimen. In these regions, tensile stresses emerge due to the shrinkage associated with the liquid-to-solid phase transformation. In contrast, negligible $\sigma_{xx}$ is observed in the remaining liquid-phase region, consistent with the assumption that the melt behaves as a viscous fluid incapable of sustaining significant shear or normal stress.

As crystallization progresses from the outer contours toward the center of the geometry, the surrounding solidified regions begin to constrain the inner liquid core. Between $t = 962$ s and $t = 1112$ s (snapshots 2-5), the top surface outline undergoes tensile stretching, while the inward-moving transformation front (e.g., at $t = 1062$ s) experiences compressive stress. This stress contrast arises from the spatial variation in crystallization kinetics—faster crystallization near the edges causes greater local shrinkage compared to the center, leading to an imbalance in deformation.

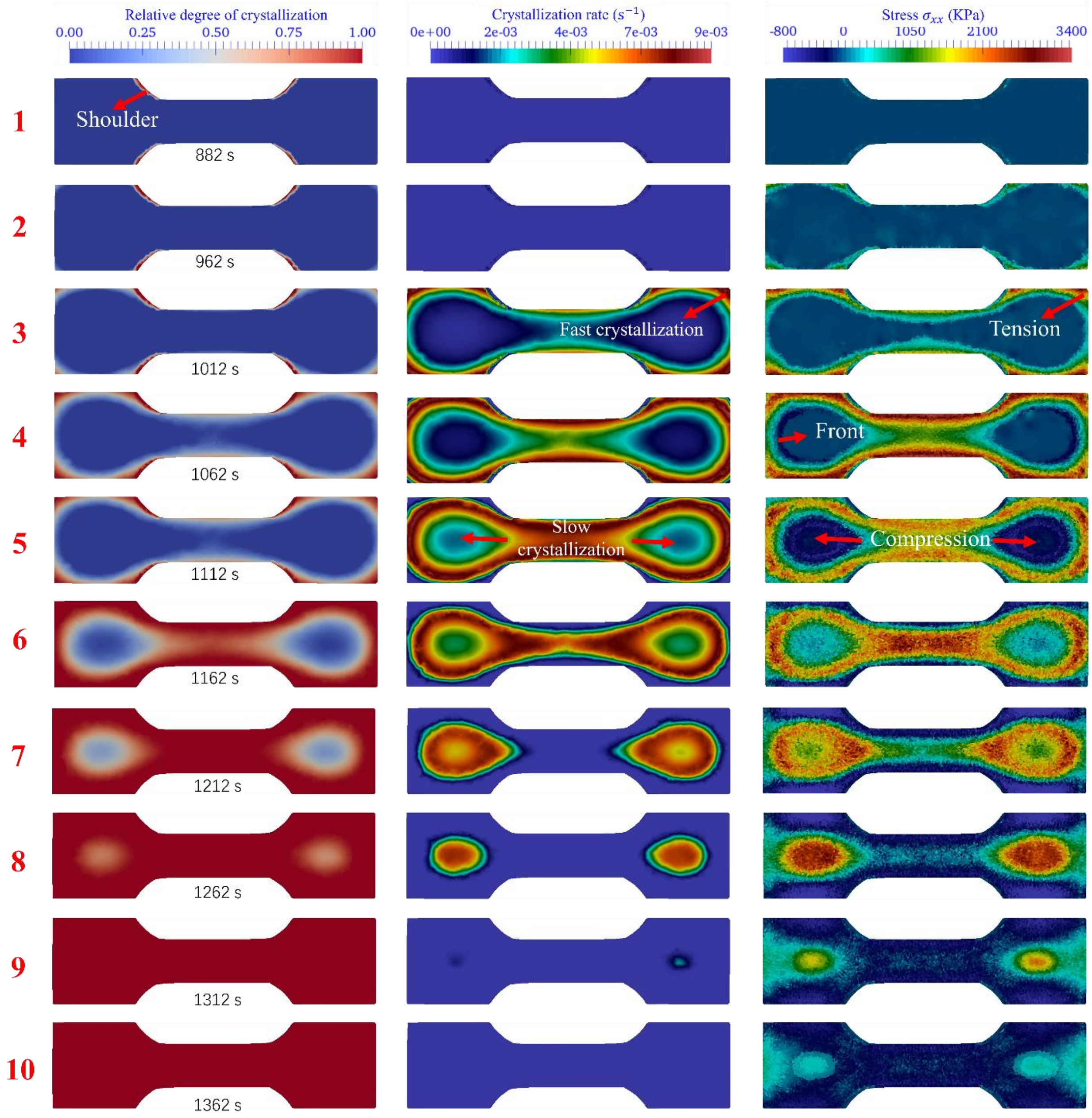

*Fig.10 Top view evolution of relative crystallinity, crystallization rate, and $\sigma_{xx}$ stress distribution over time for Case 2 (without thermal expansion). The snapshots reveal the spatial interplay between phase transformation and stress development.*

At $t = 1162$ s (snapshot 6), the outer shell is fully solidified, while the central region remains in the molten state. Continued inward material movement is now resisted by the rigidified periphery, resulting in the transformation of compressive stress in the center (seen earlier) into tensile stress. From this point on, solidification is primarily governed by central crystallization, which proceeds in a more uniform manner. As a general trend, regions undergoing rapid crystallization (higher transformation rate) tend to develop tensile stress, while regions with slower kinetics or fully crystallized domains exhibit compressive stress.

Snapshots 1-5 indicate that crystallization is concentrated around the specimen edges, coinciding with an overall increase in warpage and tensile stress on the top surface. However, after snapshot 6, the crystallization rate at the center surpasses that at the edges. This spatial shift promotes a more homogeneous crystallization process and reduces the differential shrinkage across the geometry. Consequently, warpage diminishes, and the stress state transitions from tensile at the center to compressive at the edges, following the sequence of solidification.

After complete crystallization at $t = 1362$ s, residual stress remains trapped in the structure. In the absence of external mechanical constraints, different regions of the specimen undergo self-driven deformation to relax these internal stresses. Along the central axis, the top surface tends to contract inward to relieve tensile stress, while the bottom surface expands outward. Conversely, the edge regions on the top surface, under compressive stress, tend to expand outward. This peripheral expansion becomes dominant and may account for the observed reduction in global warping after solidification is completed.

### 3.4 Effect of preheating temperature

Several studies [25,57,58] have identified the preheating temperature as a critical factor influencing the dimensional accuracy of semi-crystalline polymers processed via L-PBF. However, the detailed underlying mechanisms remain insufficiently understood. This subsection aims to investigate the effects of varying the $T_b$ and the $T_{new}$ on the displacement evolution of the fabricated part.

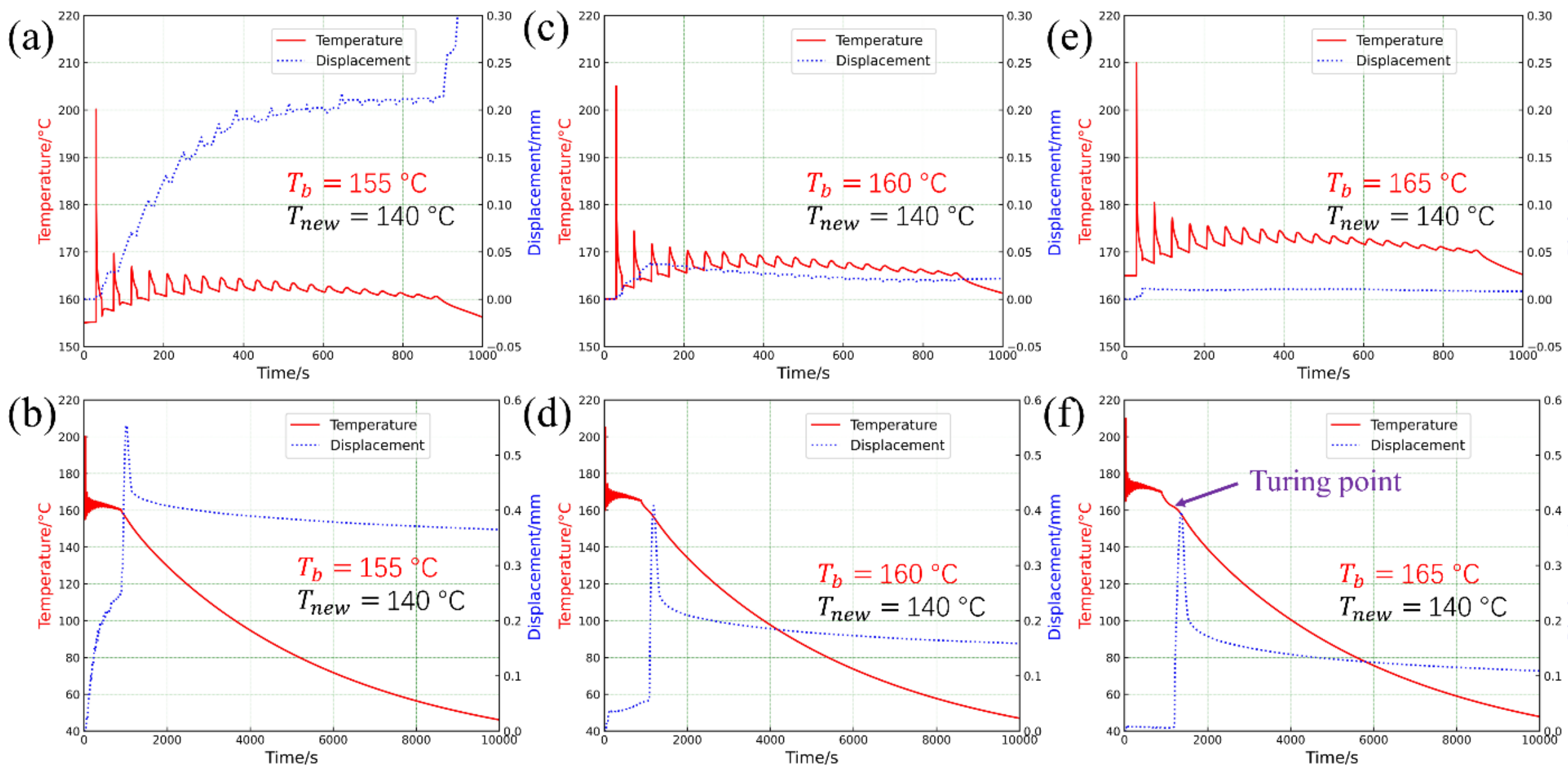

*Fig.11 Temperature and displacement evolution of the sensor 1 with preheating target temperature $T_b$ at: (a)(b)155 °C; (c)(d) 160 °C; and (e)(f) 165 °C. For these three cases, both thermal expansion and phase transformation effect are considered.*

Fig.11 presents the temperature and displacement histories recorded at sensor 1 for three different $T_b$. Fig.11(a), (c), (e) depict the evolutions during the printing stage, while Fig.11(b), (d), (f) show the corresponding curves over the entire simulation period, including the cooling stage. In all cases, $T_{new}$ is fixed at 140 °C, and both thermal expansion and crystallization-induced volumetric strain are incorporated in the thermo-mechanical model.

As indicated by the red curves, increasing $T_b$ leads to a higher peak temperature in each energy deposition cycle. More importantly, the displacement curves reveal that lower $T_b$ (e.g., $T_b$ = 155°C) are associated with sharper and larger initial displacements, while higher preheating settings significantly suppress the magnitude and rate of warpage during the printing process. This trend is clearly visible in Fig.11 (a), (c), and (e). The observed behavior is closely linked to the influence of thermal conditions on crystallization kinetics. At lower $T_b$, the molten polymer experiences steeper thermal gradients and more rapid cooling, which accelerate crystallization and intensify shrinkage-induced deformation. Conversely, an elevated $T_b$ reduces the degree of undercooling, leading to slower and more uniform crystallization. As a result, the volumetric shrinkage becomes less abrupt, effectively mitigating part warpage.

Fig.11(b), (d), and (f) further illustrate the post cooling stage. Once laser scanning is completed, due to the absence of energy supplement, the temperature decreases faster than during printing. Consequently, all cases exhibit a displacement ramping-and-recovery sequence associated with the completion of crystallization. Notably, the case with $T_b$ = 165 °C shows the largest ramping amplitude, as crystallization is largely postponed to the post cooling phase under such a high preheating condition. This behavior is corroborated by the turning point observed in the temperature curves of Fig.11(f), attributed to the release of crystallization latent heat. After cooling, the residual displacement exhibits a strong dependence on $T_b$: the lowest preheating temperature ($T_b$ = 155°C) produces the largest residual warpage, while the highest ($T_b$ = 165°C) results in the smallest. These findings highlight the critical role of preheating optimization in improving dimensional stability during polymer L-PBF.

Since crystallization during the printing process can be influenced by the temperature of the recoated powder layer, the initial temperature of the powder in the feedstock tank, denoted as $T_{new}$, may also play a non-negligible role. Physically, $T_{new}$ represents the initial thermal condition of the recoated powder before the onset of the preheating phase. To isolate and evaluate this effect, three simulation cases were performed, where $T_{new}$ was varied under an identical PID-based temperature control scheme that reported in our previous work [33].

Fig.12 illustrates the temperature and displacement histories of sensor 1 under different values of $T_{new}$, ranging from 130 °C to 150 °C in 10 °C increments. In practice, the temperature of the powder feed tank is typically maintained slightly below the $T_b$ in L-PBF systems to prevent premature material aging and undesired polymerization reactions. Consistent with this, the selected range in the simulations reflects realistic operational conditions.

Interestingly, despite a slight increase in peak displacement with lower $T_{new}$, the overall effect of the initial powder temperature on warpage is found to be negligible. This outcome can be primarily attributed to the responsiveness of the PID temperature control system, which rapidly heats the newly recoated powder layer, thereby minimizing differences in thermal history before laser exposure. As a result, the crystallization kinetics and associated deformation remain largely unaffected across the tested $T_{new}$ values.

In summary, while the preheating target temperature $T_b$ has a pronounced effect on the warpage behavior of semi-crystalline polymer parts in L-PBF, the influence of the initial recoated powder temperature $T_{new}$ appears to be minimal under active thermal control.

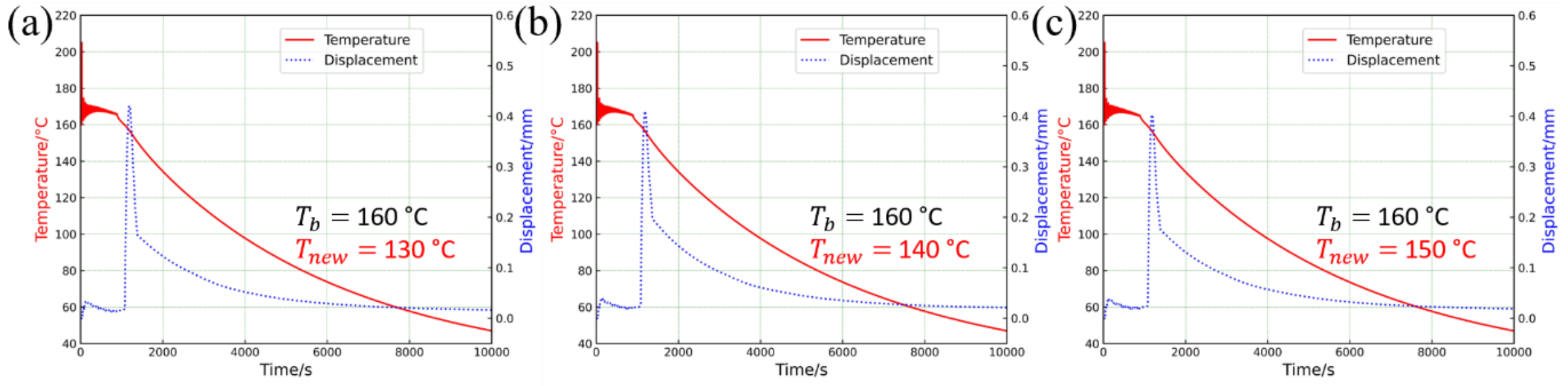

*Fig.12 Temperature and displacement evolutions with different preheating initial temperatures $T_{new}$: 130, 140, 150 °C for (a), (b), and (c) respectively.*

## 3.5 Crystallization and warpage modes with radiation boundary conditions and experimental validation

L-PBF experiments were conducted under the same processing conditions as those used in the numerical simulations. Three different preheating target temperatures ( $T_b$ =155, 160, 165 °C) and three recoated powder temperatures ( $T_{new}$ =130, 140, 150 °C) were for evaluation. For each condition, four samples were fabricated.

Fig.13(a) illustrates the spatial arrangement of the four samples (S1–S4) in each build job. Fig.13(b) and (c) present representative warpage profiles obtained from numerical simulations and experimental measurements, respectively. The experimental geometries were captured using the TM-X5120 telecentric measurement system (Keyence)[59]. Notably, upward curling of the two lateral wings is consistently observed in both the simulations and experiments.

For experimental measurement, the L-PBF manufactured dog-bone specimen is placed on a horizontal glass plate, and its profile is recorded. The top surface of the glass plate is denoted as $l_1$, and the line connecting the two curled edges of the specimen is defined as $l_2$. A reference line, $l_R$, is chosen to measure vertical distances.

The vertical distance from $l_1$ to $l_R$ is defined as $d_1$, and that from $l_2$ to $l_R$ is defined as $d_2$. The warpage of each sample, denoted as $w$, is calculated as the difference between $d_1$ and $d_2$, as illustrated in Fig.13(c):

$$w = d_1 - d_2 \quad (23)$$

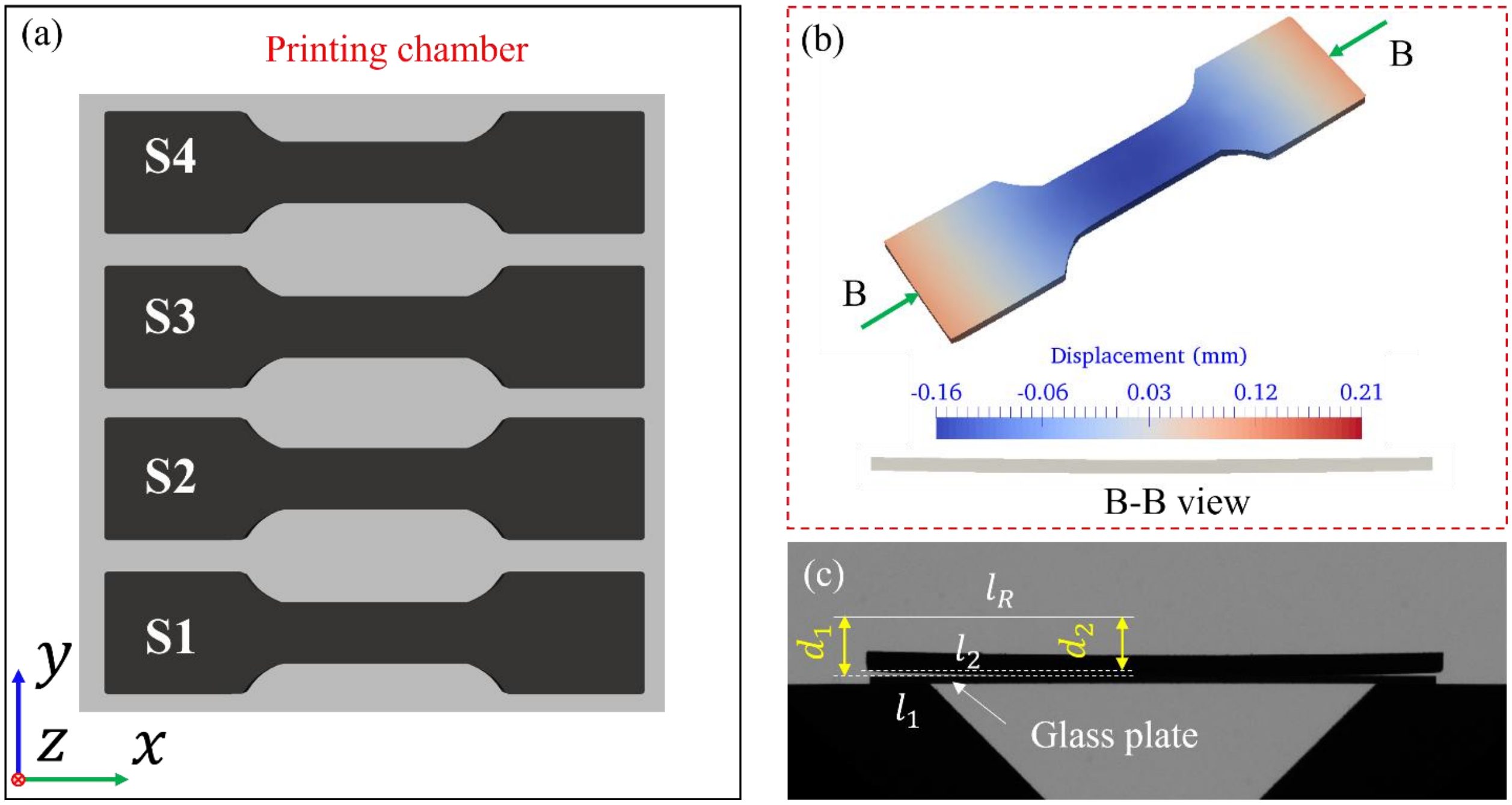

*Fig.13 (a) Position of four samples in each printing job; (b) Displacement calculated by the thermo-mechanical numerical model, and (c) Typical warping measurement by experiment.*

Table 5 summarizes the key process parameters used in the L-PBF experiments, alongside the measured warpage obtained from the optical system. For each build, samples located near the center of the powder bed (S2 and S3) consistently exhibit smaller warpage compared to those positioned near the edges (S1 and S4), likely due to more uniform thermal conditions in the central region.

In Case A, where the $T_b$ was set to 155 °C, severe warpage was observed during the early stages of printing. This excessive deformation hindered powder recoating, ultimately leading to premature build termination. In contrast, builds conducted at higher preheating temperatures of 160 °C and 165 °C (Cases B and C) were completed successfully, yielding average warpage values of 0.69 mm and 0.34 mm, respectively. Furthermore, when the preheating temperature was fixed at 165 °C, varying the initial temperature of the recoated powder $T_{new}$ between 130 °C and 150 °C (Cases C, D, and E) led to only marginal differences in the final warpage. This finding suggests that under active thermal control, the influence of $T_{new}$ on dimensional accuracy is limited compared to the decisive role of the preheating target temperatures $T_b$.

*Table 5 Parameters used for printing experiments and the resulting warping*

| Case | $T_b$ /°C | $T_{new}$ /°C | Experimental results | | | | | | | | | | | |
|---|---|---|---|---|---|---|---|---|---|---|---|---|---|---|
| | | | S1 | | | S2 | | | S3 | | | S4 | | |
| | | | $d_1$ /mm | $d_2$ /mm | $w_1$ /mm | $d_1$ /mm | $d_2$ /mm | $w_2$ /mm | $d_1$ /mm | $d_2$ /mm | $w_3$ /mm | $d_1$ /mm | $d_2$ /mm | $w_4$ /mm |
| A | 155 | 140 | - | - | huge | - | - | huge | - | - | huge | - | - | huge |
| B | 160 | 140 | 11.65 | 10.85 | 0.8 | 11.65 | 10.9 | 0.75 | 11.66 | 11.1 | 0.56 | 11.65 | 11.0 | 0.65 |
| C | 165 | 140 | 11.66 | 11.26 | 0.4 | 11.65 | 11.35 | 0.3 | 11.66 | 11.34 | 0.32 | 11.65 | 11.32 | 0.33 |
| D | 165 | 150 | 11.65 | 11.28 | 0.37 | 11.65 | 11.41 | 0.24 | 11.65 | 11.37 | 0.28 | 11.69 | 11.37 | 0.32 |
| E | 165 | 130 | 11.65 | 11.06 | 0.59 | 11.65 | 11.35 | 0.3 | 11.65 | 11.32 | 0.33 | 11.66 | 11.34 | 0.32 |

Fig.14 (a) presents a comparison between the simulated warpage results (green bars) and the corresponding experimental measurements (orange bars) under varying preheating temperatures.

In the baseline numerical simulations, a convective heat transfer coefficient of $h_b = 10\ \mathrm{W \cdot m^{-2} \cdot K^{-1}}$ is applied to the bottom surface, and $h_l = h_t = 5\ \mathrm{W \cdot m^{-2} \cdot K^{-1}}$ are assigned to the lateral and top surfaces, respectively. Under these conditions, the predicted warpage values are 0.92 mm, 0.54 mm, and 0.18 mm for $T_b$ of 155 °C, 160 °C, and 165 °C, respectively. Furthermore, as shown in simulation cases C, D and E, maintaining a preheating temperature of 165 °C while increasing the $T_{new}$ leads to a limited influence in warpage.

While the numerical trends are consistent with the experimental observations, the simulated warpage magnitudes are systematically lower. This consistent underestimation suggests that heat transfer mechanisms is not captured in the baseline model, particularly radiative heat loss from the powder surface to the surrounding air, may significantly influence the thermal field and deformation behavior. Notably, such radiation effects are rarely considered in previous L-PBF modeling literature.

To address this discrepancy, a radiative heat loss boundary condition is implemented on the top surface of the powder bed, described by:

$$\dot{q} = \sigma\epsilon(T_\Gamma^4 - T_{air}^4) \tag{24}$$

where $\sigma$ denotes the Stefan-Boltzmann constant, $\epsilon = 0.7$ is the emissivity [60], $T_\Gamma$ is any temperature on the top powder layer, and $T_{air} = 140$ °C is the air temperature in the build chamber [56].

Using $T_b,\ T_{new} = 160, 140$ °C, Fig.14 (b) compares the evolution of the relative crystallinity in the bottom three layers (L1-L3, corresponding to the same $xy$-plane position as sensor 1) of PA12 with (dashed lines) and without (solid lines) consideration of the radiation boundary condition. All crystallization curves evolve smoothly, confirming that the mesh adaptation scheme does not introduce non-physical interpolation. A distinct staircase-like pattern is observed: crystallization rate increases during the powder recoating and preheating stages, and slight reduces or stagnates during the energy deposition of the above layers, consistent with trends reported in literature [25].

Crucially, the introduction of radiative heat loss significantly accelerates the crystallization process. For instance, at approximately 300 s, the crystallinity of the third layer (violet curve) increases from 0.1 to over 0.9 when radiation is considered. This enhanced crystallization exacerbates thermally induced shrinkage. Consequently, as shown by the lavender-gray bars in Fig.14 (a), the updated warpage predictions increase substantially and show closer agreement with the experimental measurements. These findings highlight the importance of accounting for radiation effects at the powder surface when modeling thermo-mechanical behavior in polymer L-PBF processes.

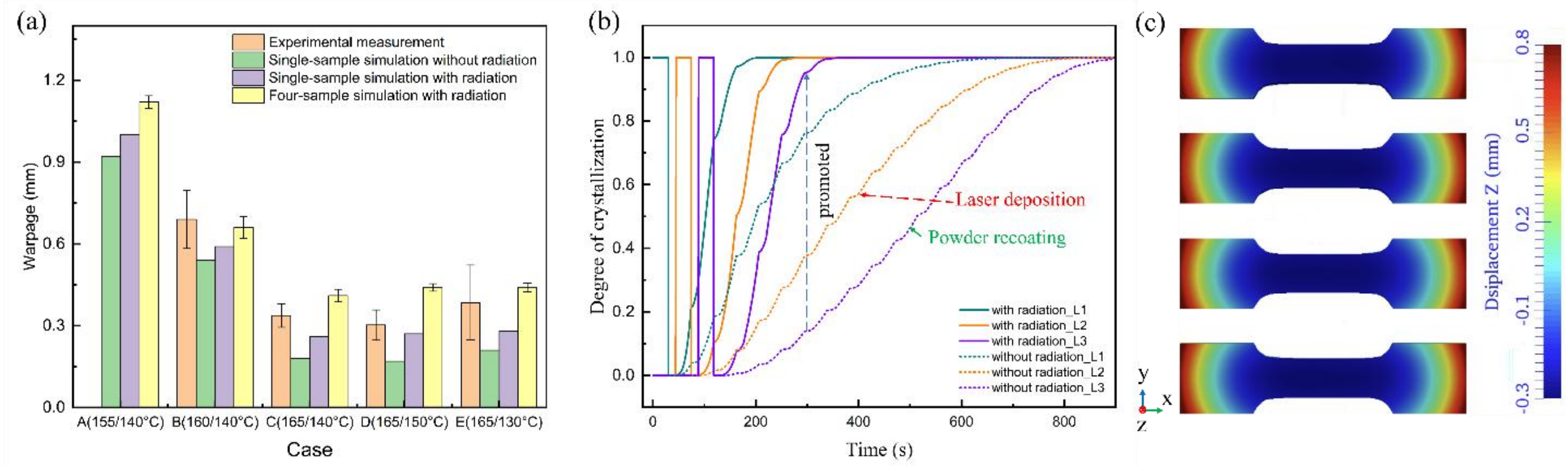

*Fig.14 (a) Comparison of simulated warpage (green bars: baseline model; lavender-gray bars: modified model with radiation boundary condition; yellow bars: simulation with four-sample configuration) and experimental measurements (orange bars) under different preheating target temperatures. (b) Evolution of the relative degree of crystallization in the bottom three layers of PA12 under a $T_b$ of 160 °C and $T_{new}$ of 140 °C, with (dashed lines) and without (solid lines) consideration of radiation heat loss. (c) final vertical displacement field of Case A ($T_b$, $T_{new}$ = 155, 140 °C) within four-sample configuration.*

To facilitate a more direct comparison with the experimental results and to better understand the origin of the warpage difference between the single- and four-dog-bone configurations, the thermo-mechanical simulations were extended to a four-dog-bone sample setup under the same set of processing parameters. As shown in Fig.3(d), a second virtual sensor (sensor 2) is placed in sample S2 of the four-dog-bone configuration at the same relative position as sensor 1 in the single-dog-bone configuration, enabling a direct comparison of the local thermal and mechanical histories between the two setups.

Fig.15 compares the temperature, vertical displacement, and phase transformation histories for the preheating condition $T_b$, $T_{new} = 165, 140$ °C. Because the scanned surface area per layer is four times larger in the four-dog-bone configuration, the total duration of each layer (including preheating, laser exposure, and dwell time) is substantially longer than in the single-dog-bone case. To account for this difference, the physical time $t$ is normalized by the total duration of one layer $t_{layer}$, yielding a normalized time $t/t_{layer}$. A minor temporal shift is applied so that the temperature peaks of the two configurations align.

On this normalized time scale, Fig.15 (a) and (b) show the temperature and vertical displacement evolutions at sensor 1 (single-dog-bone, solid lines) and sensor 2 (four-dog-bone, symbols) over the full process and during the printing stage, respectively. The first peak temperatures are comparable in both configurations, confirming that the energy input per unit area is essentially the same. However, at the end of each layer, the four-sample configuration cools to lower temperatures because its prolonged dwell time allows more energy dissipation within each layer. Under these lower temperatures sustained for a longer time, the crystallization of PA12 is significantly accelerated. This is consistent with the faster crystallization development of the four-dog-bone simulation, as illustrated in Fig.15(c). The enhanced crystallization, in turn, produces a larger volumetric shrinkage, which results in a more rapid increase in the upward vertical displacement at sensor 2. Consequently, the accumulated layer-wise shrinkage is stronger in the four-dog-bone configuration, providing a mechanistic explanation for the larger overall warpage predicted by this model compared with the single-dog-bone case.

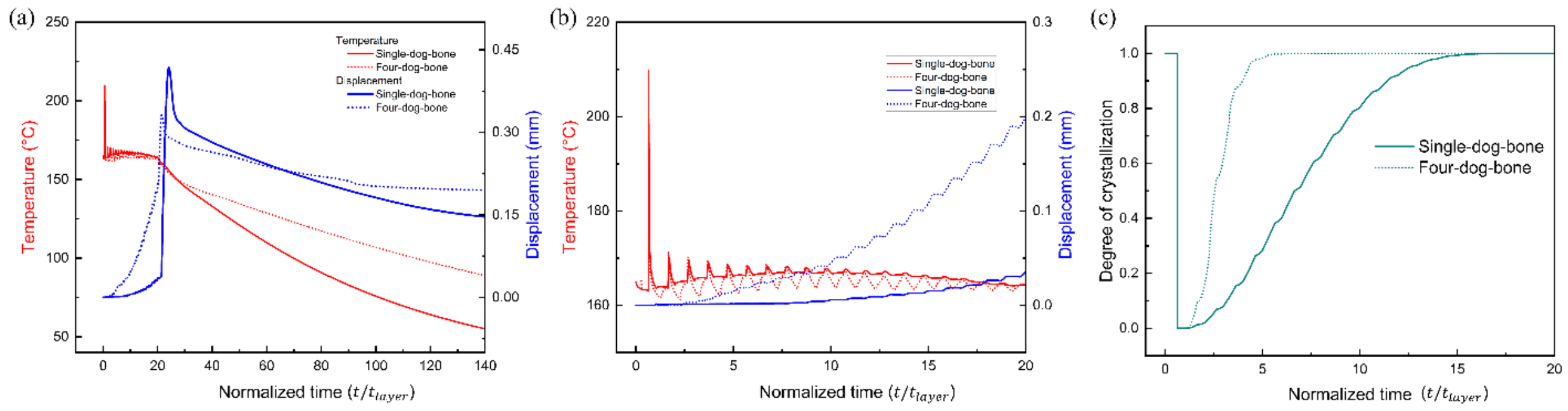

*Fig.15 Comparison of single-dog-bone and four-dog-bone simulations at the sensor locations: (a) temperature and vertical displacement over the full normalized time history; (b) temperature and vertical displacement during the printing stage; (c) evolution of the degree of crystallinity captured by sensor 1 (single-dog-bone, solid lines) and sensor 2 (four-dog-bone, dots).*

Four-sample simulations with radiation (yellow bars) are also included in Fig.14(a). Compared with the single-sample simulations, the predicted warpage systematically increases across all cases when the four-dog-bone configuration is used, consistent with the enhanced crystallization and accumulated shrinkage discussed above in connection with Fig.15. Under these conditions, the simulated warpage in the four-sample configuration tends to exceed the experimental measurements. For illustration, Fig.14(c) presents the final vertical displacement field of Case A ($T_b$, $T_{new} = 150, 140$ °C) obtained from the four-sample configuration. The variation in warpage among the four samples (corresponding to the experimental arrangement S1–S4) is smaller than that observed experimentally, suggesting that the current thermal boundary conditions could be further refined and underscoring the importance of accurately characterizing heat exchange behavior in future work.

Although the four-sample configuration provides the closest representation of the experimental setup, the single-sample simulation with an appropriate radiation boundary condition is still able to capture the general warpage trends and thus serves as a computationally efficient alternative for parametric studies. The average computational times for the single- and four-sample simulations are approximately 36 h and 68 h, respectively, reflecting the trade-off between computational cost and experimental fidelity.

## Conclusion

In this work, a comprehensive thermo-mechanical model was developed to investigate the deformation behavior of semi-crystalline polymer PA12 during the L-PBF process. The model integrates transient thermal history, crystallization kinetics, latent heat effects, and temperature-dependent viscoelasticity within a dual-mesh simulation framework.

Key conclusions are summarized as follows:

1. Phase transformation induced shrinkage dominates the warpage evolution in PA12 L-PBF. Compared to thermal expansion, crystallization-induced volume change contributes more significantly to layer-wise deformation.
2. The mechanical boundary condition plays a critical role in determining final part distortion. Modeling the bottom powder bed as a semi-supportive viscous layer provides a more realistic constraint, enabling upward warpage that is consistent with experimental observations.
3. The inclusion of radiative heat loss at the powder surface is essential for accurate warpage prediction. For the single-dog-bone configuration, the average warpage prediction error decreases from 39.4% (without radiation) to 18.6% (with radiation). For the four-sample configuration with radiation, the average prediction error is 21.4%, which is 18% lower than in the single-sample case without radiation, suggesting that radiation-driven enhancement of crystallization is a key factor in narrowing the discrepancy between numerical and experimental results.
4. Parametric studies of preheating strategies reveal that increasing the target bed temperature effectively suppresses warpage by reducing the degree of undercooling and crystallization rate, whereas variations in the initial recoated powder temperature have minimal impact under active thermal control.

The model predictions show good agreement with experimental measurements, validating its potential as a predictive tool for optimizing thermal management and mechanical performance in polymer-based L-PBF. Future work will extend this modeling framework to account for scanning strategy, and in-situ monitoring strategies for enhanced process control.

## Acknowledgments

This work was financially supported by the China Scholarship Council (Grant No. [2020]08440261). Wei Zhu acknowledges the support from the National Natural Science Foundation of China (Grant No. 52205355) and the Opening Project of the Guangdong Provincial Key Laboratory of Technique and Equipment for Macromolecular Advanced Manufacturing (Grant No. 2024kfkt05).